\documentclass[preprint,3p, times]{elsarticle}
\usepackage{amsfonts,amsmath,amsthm}
\usepackage{mathrsfs}
\usepackage[hidelinks]{hyperref}
\usepackage{graphicx}
\usepackage{epstopdf}
\usepackage{caption}
\usepackage{subcaption}
\usepackage{enumitem}
\usepackage{color}

\newcommand{\vect}[1]{\mathbf{#1}}                    
\newcommand{\vects}[1]{\pmb{#1}}                      
\newcommand{\vectNF}[1]{\hat{\mathbf{#1}}}            
\newcommand{\mat}[1]{\mathbb{#1}}                     
\newcommand{\varNF}[1]{\tilde{#1}}                    

\providecommand{\abs}[1]{\lvert#1\rvert}              
\newcommand{\opt}[1]{\pmb{\mathcal{#1}}^{\dag}}       
\newcommand{\opts}[1]{\mathcal{#1}^{\dag}}            
\newcommand{\normp}[1]{\left(#1\right)}               
\newcommand{\normb}[1]{\left[#1\right]}               

\newcommand{\onehalf}{\frac{1}{2}}
\newcommand{\dudt}[1]{\frac{\partial #1}{\partial t}} 
\newcommand{\dudx}[1]{\frac{\partial #1}{\partial x}} 

\newcommand{\intx}[1]{\left<#1\right>_x}               
\newcommand{\intt}[1]{\left<#1\right>_t}               
\newcommand{\intxt}[1]{\left<#1\right>_{xt}}           

\newcommand{\refEq}[1]{Eq. (\ref{#1})}
\newcommand{\refFig}[1]{Figure \ref{#1}}
\newcommand{\refTab}[1]{Table \ref{#1}}

\begin{document}
\begin{frontmatter}
\title{Assessment of continuous and discrete adjoint method for sensitivity analysis in two-phase flow simulations}
\author{Guojun Hu\corref{corAuthor}}
\ead{ghu3@illinois.edu}
\author{Tomasz Kozlowski\corref{}}
\ead{txk@illinois.edu}
\address{Department of Nuclear, Plasma, and Radiological Engineering, University of Illinois at Urbana-Champaign\\
	Talbot Laboratory 104 S Wright St, Urbana IL, 61801, United States}

\cortext[corAuthor]{Corresponding author}

\begin{abstract}
	Verification, validation and uncertainty quantification (VVUQ) have become a common practice in thermal-hydraulics analysis. An important step in the uncertainty analysis is the sensitivity analysis of various uncertainty input parameters. The efficient method for computing the sensitivities is the adjoint method. The cost of solving an adjoint equation is comparable to the cost of solving the governing equation. Once the adjoint solution is obtained, the sensitivities to any number of parameters can be obtained with little effort. There are two methods to develop the adjoint equations: continuous method and discrete method. In the continuous method, the control theory is applied to the forward governing equation and produces an analytical partial differential equation for solving the adjoint variable; in the discrete method, the control theory is applied to the discrete form of the forward governing equation and produces a linear system of equations for solving the adjoint variable. In this article, an adjoint sensitivity analysis framework is developed using both the continuous and discrete methods. These two methods are assessed with the one transient test case based on the BFBT benchmark. Adjoint sensitivities from both methods are verified by sensitivities given by the perturbation method. Adjoint sensitivities from both methods are physically reasonable and match each. The sensitivities obtained with discrete method is found to be more accurate than the sensitivities from the continuous method. The continuous method is computationally more efficient than the discrete method because of the analytical coefficient matrices and vectors. However, difficulties are observed in solving the continuous adjoint equation for cases where the adjoint equation contains sharp discontinuities in the source terms; in such cases, the continuous method is not as robust as the discrete adjoint method.
\end{abstract}


\end{frontmatter}

\section{Introduction}
In recent years, verification, validation and uncertainty quantification (VVUQ) have become a common practice in thermal-hydraulics analysis. In general, these activities deal with propagation of uncertainties in computer code simulations, e.g., through system analysis codes. An important step in uncertainty analysis is the sensitivity analysis of various uncertain input parameters. The common method to calculate sensitivity includes variance-based and regression-based methods. However, these methods require solving the system of interest multiple times, sometimes 100s of times, which is computationally expensive. An alternative method for calculating sensitivities is the adjoint method. The cost of solving an adjoint equation is comparable to the cost of solving the governing equation (forward equation, e.g. the two-phase two-fluid model). However, once the adjoint solution is obtained, the sensitivity to an arbitrary number of parameters can be calculated at the same time.

Adjoint equations are not rare throughout mathematics and its applications. There is a long history of using the adjoint method in optimal control theory. The use of adjoint method for computing sensitivities came up in the nuclear industry in the 1940s for the estimation of first eigenvalues of the transport equation. Later, the adjoint method became popular in computational fluid dynamics field \cite{Marchuk1995, Giles2000}. Within the field of aeronautical computational fluid dynamics, the use of adjoint method has been seen in \cite{Jameson1988, Jameson1994, Jameson1998, Nadarajah2000}. Adjoint problems arise naturally in the formulation of methods for optimal aerodynamic design and optimal error control \cite{Giles1998,Giles2000, Giles2001, Giles2003}. Adjoint solution provides the linear sensitivities of an objective function (e.g. lift force and drag force) to a number of design variables. These sensitivities can then be used to drive an optimization procedure. In a sequence of papers, Jameson developed the adjoint method for the potential flow, the Euler equation, and the Navier-Stokes equation \cite{Jameson1988, Jameson1994, Jameson1998, Nadarajah2000}. The application of the adjoint method to optimal aerodynamic design is very successful. However, to the author's best knowledge, successful adjoint sensitivity analysis to two-phase flow problems is rare. Cacuci performed an adjoint sensitivity analysis to two-phase flow problems using the RELAP5/MOD3.2 numerical discretization \cite{Cacuci1982, Cacuci2000a, Cacuci2000b}. An application of Cacuci's method was illustrated by \cite{petruzzi2008development}, where the method was applied to the blow-down of a gas from a pressurized vessel. 

There are two methods to develop the adjoint equations: continuous and discrete \cite{nadarajah2003discrete}. In the continuous method, the control theory is applied to the governing equation through a variation analysis. The variation of the governing equation and the response function with respect to the primitive variables (e.g. pressure and velocity) and input parameters (e.g. boundary conditions) are combined through the use of Lagrange multipliers (adjoint variables). The continuous adjoint equation and its boundary conditions are specified by collecting the terms associated with the variation of the primitive variables; while the sensitivity is specified by the terms associated with the variation of the input parameters. The continuous adjoint equation is then discretized to obtain numerical solutions.  In the discrete method, the control theory is applied directly to the discrete form of the governing equation (e.g. the residual vector given by the discretization). The discrete adjoint equation is specified by collecting terms associated with the variation of the primitive variables; while the sensitivity is specified by collecting terms associated with the variation of the input parameters. The discrete adjoint equation is usually given in the form of a system of linear equations. 

The main advantage of the continuous adjoint method is that it provides a continuous adjoint equation in an analytical form, which does not depend on the solver for the forward equation and can be solved separately. This analytical form helps researchers study and understand the nature of the equation. The main disadvantage of the continuous method is that it requires one to develop an appropriate algorithm to discretize the adjoint equation and its boundary conditions. The development, verification, and validation of the adjoint code is complicated due to the lack of suitable benchmark test cases. The main advantage of the discrete adjoint method is that the code development is a straightforward process because it is closely related the forward solver. The variation of the discrete forward equations produces the coefficient matrices and vectors for the discrete adjoint equation. These coefficient matrices and vectors can be obtained either manually or by automatic differentiation. The robustness and accuracy of the adjoint code is the same as that of the forward code because the eigenvalues of the two systems are the same. The main disadvantage of the discrete method is the extra complexity in obtaining coefficient matrices/vectors manually or the extra cost in obtaining coefficient matrices/vectors with automatic differentiation. 

In previous work \cite{Hu2018PhdThesis, Hu2018CAdjoint, Hu2018DAdjoint}, the author applied separately the continuous and adjoint methods to sensitivity analysis in steady-state two-phase flow simulations. Both methods were found to work well. However, these two methods were based on different forward solver and were assessed with different test problems. This article will assess both methods within the same framework for a consistent comparison. In additions, the previous studies were limited to steady-state problems. Since the transient behavior of the two-phase system is important in reactor safety analysis, this article will assess the application of both continuous and adjoint methods to sensitivity analysis in transient two-phase flow simulations. 

This article is organized in the following way. Section \ref{sec-two} presents briefly the forward numerical scheme for solving the two-phase two-fluid model equation. Section \ref{sec-three} presents the adjoint sensitivity analysis framework for both continuous and discrete adjoint method for transient two-phase flow simulations. Section \ref{sec-four} presents the numerical tests for verification and assessment of both continuous and discrete adjoint methods. The faucet flow will be used to assess the application of both methods to steady-state problems. One transient test in BFBT benchmark will be used to assess the application of both methods to realistic transient two-phase flow problems. Section \ref{sec-five} presents the conclusion and the future work of the current adjoint sensitivity analysis framework.

\section{Forward solver}\label{sec-two}
\subsection{Two-phase two-fluid model}\label{sec-two-p1}
For 1D problems, the basic two-phase two-fluid six-equation model without any differential closure correlations \cite{ishii2010thermo} can be written in a vector form as
\begin{equation}\label{forward-governing-equation}
    \dudt{\vect{U}} + \dudx{\vect{F}} + \vect{P}_{ix}\dudx{\alpha_g} + \vect{P}_{it}\dudt{\alpha_g} = \vect{S}
\end{equation}
where $\vect{U}$ is the vector of conservative variables, $\vect{F}$ is the vector of flux variables, $\vect{P}_{ix}$ and $\vect{P}_{it}$ are the vectors related to the partial derivatives of the void fraction, and $\vect{S}$ is the vector of source terms. They are defined as
 \begin{equation}\label{forward-governing-equation-vector-def}
    \vect{U} \equiv \begin{pmatrix}
                \alpha_l\rho_l \\
                \alpha_l\rho_l u_l \\
                \alpha_l\rho_l E_l\\
                \alpha_g\rho_g \\
                \alpha_g\rho_g u_g \\
                \alpha_g\rho_g E_g\\
                \end{pmatrix},
     \vect{F} \equiv \begin{pmatrix}
                \alpha_l\rho_l u_l \\
                \alpha_l\rho_l u_l^2 + \alpha_l p \\
                \alpha_l\rho_l H_l u_l\\
                \alpha_g\rho_g u_g\\
                \alpha_g\rho_g u_g^2 + \alpha_g p \\
                \alpha_g\rho_g H_g u_g\\
                \end{pmatrix},
      \vect{W} \equiv \begin{pmatrix}
                \alpha_g \\
                p \\
                T_l\\
                T_g \\
                u_l \\
                u_g\\
                \end{pmatrix}, \vect{P}_{ix} \equiv \begin{pmatrix}
                        0 \\
                        p \\
                        0 \\
                        0 \\
                       -p \\
                        0\\
                    \end{pmatrix},
     \vect{P}_{it} \equiv \begin{pmatrix}
                         0 \\
                         0 \\
                        -p \\
                         0 \\
                         0 \\
                         p\\
                    \end{pmatrix}
\end{equation}
Let the subscript $k=l,g$ denote the liquid phase and gas phase, respectively. The variables $\normp{\alpha_k, \rho_k, u_k, e_k}$ denote the volume fraction, the density, the velocity, and the specific internal energy of $k$-phase. The summation of phasic volume fraction should be one, i.e. $\alpha_l + \alpha_g = 1$.  $p$ is the pressure of two phases. $E_k = e_k + u_k^2/2$ and $H_k = e_k + p/\rho_k + u_k^2/2$ are the phasic specific total energy and specific total enthalpy.

An appropriate Equation of State (EOS) is required to close the system. For many practical problems in the nuclear thermal-hydraulics analysis, the temperature of two phases are required to model the source terms. In such a case, a useful EOS is given by specifying the Gibbs free energy  as a function of pressure and temperature $T_k$, i.e.
\begin{equation}
\mathfrak{g}_k = \mathfrak{g}_k(T_k,p), \text{  for  }k=l,g
\end{equation}
After specifying the specific Gibbs free energy, the phasic density and specific internal energy are obtained from the partial derivatives of the specific Gibbs free energy. The details about specifying the EOS through the specific Gibbs free energy are referred to \cite{iapws1998, Hu2018PhdThesis}.

\subsection{Numerical scheme}\label{sec-two-p2}
\begin{figure}[!htb]
  \centering
  \includegraphics[width=0.45\textwidth]{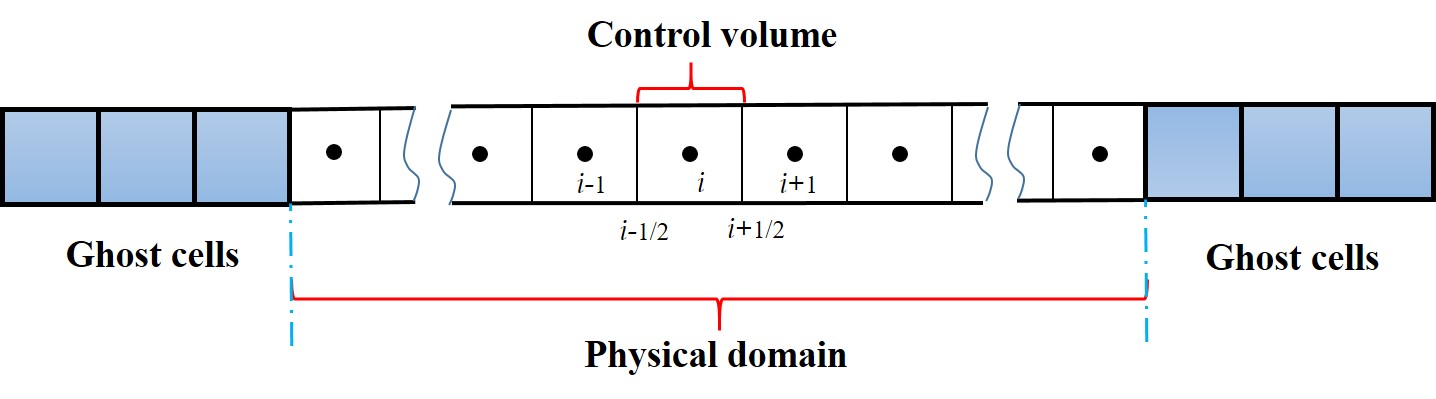}\\
  \caption{Schematic of the 1D spatial discretization}\label{schematic-mesh}
\end{figure}

For 1D problems, the spatial discretization is shown schematically in \refFig{schematic-mesh}. The physical domain is divided into $N$ cells. The cell center is denoted with an index $i$ and the cell boundaries are denoted with $i\pm 1/2$, for $i=1,\cdots, N$. All unknown variables are solved in the cell center (co-located mesh). On each side of the physical domain, ghosts cells are used to deal with boundary conditions.

The two-phase two-fluid model is solved with the backward Euler method, which gives a fully implicit scheme
\begin{equation}\label{backward-euler-fully-implicit}
    \vect{U}_i^{n+1}-\vect{U}_i^{n} + \vect{P}_{it,i}^{n+1}\normp{\alpha_{g,i}^{n+1}-\alpha_{g,i}^{n}}=  \Delta t \opt{L}\normp{\vect{U}_i^{n+1}}
\end{equation}
where $\opt{L}$ is an operator representing the spatial differential operators and the source terms,
\begin{equation}\label{backward-euler-operator}
      \opt{L}\normp{\vect{U}_i}= - \frac{\vectNF{F}_{i+1/2} -\vectNF{F}_{i-1/2}}{\Delta x}-\vect{P}_{ix,i}\frac{\alpha_{g,i+1}-\alpha_{g,i-1}}{2\Delta x} + \vect{S}_i
\end{equation}
where $\vectNF{F}_{i+1/2}$ and $\vectNF{F}_{i-1/2}$ are low-order numerical fluxes at cell boundaries. More details of the numerical scheme can be seen in \cite{Hu2018Implicit}.

A first-order Roe-type numerical flux is constructed following the Roe-Pike \cite{toro2013, glaister1988approximate} method. Let $\mat{A}_c$ be the Jacobian matrix defined as $\mat{A}_c \equiv \partial \vect{F}/\partial \vect{U}$. The subscript $c$ denotes that the Jacobian matrix and eigenvalues/eigenvectors are obtained with the conservative part of the equation \cite{Hu2017Riemann, Hu2018PhdThesis}. Let $\lambda_{m}$ and $\vect{K}_{m}$, for $m = 1, \cdots, 6$ be the eigenvalues and right eigenvectors of the matrix $\mat{A}_c$. Let $Q(z)$ be a scalar function defined as
\begin{equation}
    Q(z) = \left\{
     \begin{array}{l}
       \frac{1}{2}\normp{\frac{z^2}{\delta} + \delta}, \quad \abs{z} < \delta \\
       \abs{z}, \quad \quad \quad \quad \abs{z} \geq \delta
     \end{array}\right.
\end{equation}
where $\delta$ is the coefficient for the addition of numerical viscosity term, which is set at 0.125 as was used by Yee \cite{Yee1985}.  The Roe-type numerical flux is constructed by
\begin{equation}\label{R-L-Flux}
    \vectNF{F}_{i+1/2} = \onehalf\normp{\vect{F}_i +\vect{F}_{i+1}} - \onehalf \sum_{m=1}^{6}\varNF{\omega}_{m, i+1/2} Q\normp{\varNF{\lambda}_{m, i+1/2}}\varNF{\vect{K}}_{m, i+1/2}
\end{equation}
where $\omega_{m, i+1/2}$ is the wave strength when projecting the conservative vectors to the characteristic space. The Jacobian matrix, approximate eigenvalues, and eigenvectors are given in the Appendix A for reference. Complete details about the approximate eigenvalues/eigenvectors and the average state are referred to \cite{Hu2017Riemann, Hu2018PhdThesis}. 

\refEq{backward-euler-fully-implicit} forms a set of algebraic nonlinear equations, which are solved  with the JFNK method. \refEq{backward-euler-fully-implicit} can be generalized as
\begin{equation}
    \vect{G}(\vect{W}) = \vect{0}
\end{equation}
where $\vect{G}$ denotes the global residual function of \refEq{backward-euler-fully-implicit} and $\vect{W}$ is the vector of unknown primitive variables. The JFNK method is based on the Newton's method, which solves the nonlinear algebraic equations iteratively
\begin{align}
    \mat{J}^{m}\delta \vect{W}^{m} &= - \vect{G}(\vect{W}^{m}) \label{JFNK-linear-eq} \\
    \vect{W}^{m+1} &= \vect{W}^{m} + \delta \vect{W}^{m}\label{JFNK-update}
\end{align}
where $m$ denotes the $m$-th step of the iteration. The iteration starts with an initial guess of $\vect{W}$ which is usually taken from the old time step. $\mat{J}^{m}$ is the Jacobian matrix define as
\begin{equation}
    \mat{J}^{m} \equiv \normp{\frac{\partial \vect{G}}{\partial \vect{W}}}^{m}
\end{equation}
In the JFNK scheme, the linear equation \refEq{JFNK-linear-eq} is solved with the Krylov subspace method. The essential idea of the JFNK method is that the Krylov method requires only the matrix-vector product and the explicit form of the Jacobian matrix could be avoided. The matrix-vector project is approximated with
\begin{equation}
    \mat{J}^{m}\vect{v} \approx \frac{\vect{G}(\vect{W}^{m} + \epsilon \vect{v})-\vect{G}(\vect{W}^{m})}{\epsilon}
\end{equation}
where $\vect{v}$ is the Krylov vector and $\epsilon$ is a small parameter. In this article, the JFNK method is implemented with the scientific computational toolkit PETSc \cite{balay2016petsc}.

\section{Adjoint sensitivity analysis}\label{sec-three}
\subsection{General framework}
Let $\opt{G}$ be the operator that represents the governing equation of the forward problem, e.g. the two-phase two-fluid equation. Let $\vect{W}$ be the vector of physical variables. For the forward problem, there are usually a few parameters ($\vects{\omega}$) that affect the solution, e.g. the physical model parameters and boundary conditions. Suppose the governing equation is written as
\begin{equation}
\opt{G}\normp{\vect{W}, \vects{\omega}} = \vect{0}
\end{equation}
Let $\opts{R}$ be the operator that measures the quantity of interest (response function), e.g. void fraction.  The response could be expressed as
\begin{equation}
R = \opts{R}\normp{\vect{W}, \vects{\omega}}
\end{equation}
In the following discussion, vectors and matrices are defined in a way such that the multiplications shown in the following equations are the inner product.

Let $\delta$ be the variation operator. A perturbation in the parameter of interest, $\delta\vects{\omega}$, will cause
\begin{equation}\label{variation-in-forward-ge}
\normp{\frac{\partial \opt{G}}{\partial \vect{W}}}_{\vects{\omega}}\delta\vect{W} + \normp{\frac{\partial \opt{G}}{\partial \vects{\omega}}}_{\vect{W}}\delta\vects{\omega} = \vect{0}
\end{equation}
\begin{equation}\label{variation-in-response}
\delta R = \normp{\frac{\partial \opts{R}}{\partial \vect{W}}}_{\vects{\omega}}\delta\vect{W} + \normp{\frac{\partial \opts{R}}{\partial \vects{\omega}}}_{\vect{W}}\delta\vects{\omega}
\end{equation}
Note that $\delta R/\delta\vects{\omega}$ is the sensitivity we would like to quantify. Solving for $\delta \vect{W}$ with \refEq{variation-in-forward-ge} is usually expensive because it requires solving the forward problems. The idea of the adjoint method is to remove the dependency of $\delta R$ on $\delta \vect{W}$ by combining \refEq{variation-in-forward-ge} and \refEq{variation-in-response} using Lagrange multipliers.

For brevity, the subscripts $\vects{\omega}$ and $\vect{W}$ in the partial derivatives are dropped. Let $\vects{\phi}$ be a vector of free variables which represents the vector of Lagrange multipliers.  Multiplying the transpose of the Lagrange multiplier to \refEq{variation-in-forward-ge} gives
\begin{equation}\label{adjoint-multiply-variation-ge}
\vects{\phi}^{T}\normb{\frac{\partial \opt{G}}{\partial \vect{W}}\delta\vect{W} + \frac{\partial \opt{G}}{\partial \vects{\omega}}\delta\vects{\omega}} = 0
\end{equation}
where the superscript $T$ denotes the transpose operator. Since the right-hand side of \refEq{adjoint-multiply-variation-ge} is zero, \refEq{adjoint-multiply-variation-ge} can be subtracted from \refEq{variation-in-response} without changing the value of $\delta R$, i.e.
\begin{equation}\label{dR-minus-adjoint-multiply-variation-ge}
\delta R = \frac{\partial \opts{R}}{\partial \vect{W}}\delta\vect{W} + \frac{\partial \opts{R}}{\partial \vects{\omega}}\delta\vects{\omega} - \vects{\phi}^{T}\normb{\frac{\partial \opt{G}}{\partial \vect{W}}\delta\vect{W} + \frac{\partial \opt{G}}{\partial \vects{\omega}}\delta\vects{\omega}}
\end{equation}
\refEq{dR-minus-adjoint-multiply-variation-ge} is rewritten as
\begin{equation}\label{adjoint:response-with-W}
\delta R = \normp{\frac{\partial \opts{R}}{\partial \vect{W}}-\vects{\phi}^{T}\frac{\partial \opt{G}}{\partial \vect{W}}}\delta\vect{W} + \normp{\frac{\partial \opts{R}}{\partial \vects{\omega}}- \vects{\phi}^{T}\frac{\partial \opt{G}}{\partial \vects{\omega}}}\delta\vects{\omega}
\end{equation}
Because the Lagrange multiplier $\vects{\phi}$ is a vector of free variables, it can be chosen in a way such that
\begin{equation}\label{transpose-general-adjoint-equation}
\frac{\partial \opts{R}}{\partial \vect{W}}-\vects{\phi}^{T}\frac{\partial \opt{G}}{\partial \vect{W}} = \vect{0}
\end{equation}
The so-called adjoint equation is obtained by taking the transpose of \refEq{transpose-general-adjoint-equation}, i.e.
\begin{equation}\label{general-adjoint-equation}
\normp{\frac{\partial \opt{G}}{\partial \vect{W}}}^{T}\vects{\phi} - \normp{\frac{\partial \opts{R}}{\partial \vect{W}}}^{T} = \vect{0}
\end{equation}
The Lagrange multiplier $\vects{\phi}$ given by \refEq{general-adjoint-equation} is the so-called adjoint variable. With \refEq{general-adjoint-equation}, the sensitivity is obtained with
\begin{equation}\label{general-adjoint-sensitivity}
\frac{\mathrm{d}R}{\mathrm{d}\vects{\omega}} =  \normp{\frac{\partial \opts{R}}{\partial \vects{\omega}}- \vects{\phi}^{T}\frac{\partial \opt{G}}{\partial \vects{\omega}}}
\end{equation}
Note that if the response function does not depend explicitly on $\vects{\omega}$, then $\partial\opts{R}/\partial\vects{\omega}$ can be removed from \refEq{general-adjoint-sensitivity}. The advantage of \refEq{general-adjoint-sensitivity} is that it is independent of $\delta\vect{W}$, which means that the sensitivity of the response to an arbitrary number of parameters can be determined without the need for additional forward calculations.

\subsection{Application of continuous adjoint method to two-fluid model}
\subsubsection{Continuous adjoint equation}
The adjoint equation \refEq{general-adjoint-equation} and the response equation \refEq{general-adjoint-sensitivity} are problem dependent. The continuous adjoint method works with the PDE of the forward problem, i.e. \refEq{forward-governing-equation}. Note that the $\vect{P}_{it}\partial \alpha_g/\partial t$ term term can be handled by the matrix $\mat{A}_{it} = \vect{P}_{it} \partial\alpha_g/\partial\vect{U}$, i.e.
\begin{equation}
\normp{\mat{I} + \mat{A}_{it}}\dudt{\vect{U}} + \dudx{\vect{F}} + \vect{P}_{ix}\dudx{\alpha_g} -\vect{S}=\vect{0}
\end{equation}
Because $\normp{\mat{I} + \mat{A}_{it}}$ and its inverse can be obtained analytically, its effect can be considered later if interested. The derivation is started with
\begin{equation}\label{tr-forward-governing-equation}
\dudt{\vect{U}} + \dudx{\vect{F}} + \vect{P}_{ix}\dudx{\alpha_g} -\vect{S}=\vect{0}
\end{equation}

Let $\intx{*}$ denote the integration in space, $\intt{*}$ denote the integration in time, and $\intxt{*}$ denote the double integration in space and time. The integration in time is from $t_0$ to time $t_1$ and the integration in space is from $x_0$ to $x_1$. In this article, the response function is generalized as
\begin{equation}
R = \intxt{q\normp{x, t}}
\end{equation}
where $q$ is the quantity of interest which is a function of the primitive variables ($\vect{W}$). Applying the variation operator to the response function gives,
\begin{equation}\label{adjoint-variation-R}
\delta R = \intxt{\frac{\partial q}{\partial \vect{W}}\delta\vect{W}}=\intxt{-\vect{Q}\delta \vect{W}}, \text{  with  } \vect{Q}\equiv -\frac{\partial q}{\partial \vect{W}}
\end{equation}

Let $\vects{\phi}$ be the Lagrange multiplier (the adjoint variable). Multiplying $\vects{\phi}^{T}$ to the variation of \refEq{tr-forward-governing-equation} and integrating the product over space and time gives
\begin{equation}\label{variation-in-forward-ge4}\begin{split}
-\intxt{\normp{\dudt{\vects{\phi}^{T}}\mat{A}_0 + \dudx{\vects{\phi}^{T}}\mat{A}_{1} +\vects{\phi}^{T}\mat{A}_2}\delta\vect{W}} &-\intxt{\vects{\phi}^{T}\frac{\partial \vect{S}}{\partial \vects{\omega}}\delta\vects{\omega}}\\
&\intx{\vects{\phi}^{T}\mat{A}_0\delta\vect{W}|^{t_1} - \vects{\phi}^{T}\mat{A}_0\delta\vect{W}|^{t_0}} + \intt{\vects{\phi}^{T}\mat{A}_1\delta\vect{W}|^{x_1} - \vects{\phi}^{T}\mat{A}_1\delta\vect{W}|^{x_0}} = 0
\end{split}\end{equation}
where the following matrices are defined
\begin{equation}\begin{split}
\mat{A}_{0} &=\frac{\partial \vect{U}}{\partial \vect{W}}, \mat{A}_{1} = \frac{\partial \vect{F}}{\partial \vect{W}}+\vect{P}_{ix}\frac{\partial \alpha_g}{\partial \vect{W}}\\
\mat{A}_2 &= \dudx{\vect{P}_{ix}}\frac{\partial \alpha_g}{\partial \vect{W}}- \frac{\partial \vect{P}_{ix}}{\partial \vect{W}}\dudx{\alpha_g} + \frac{\partial \vect{S}}{\partial \vect{W}}
\end{split}\end{equation}
Note that the effect of $\vect{P}_{it}\partial \alpha_g/\partial t$ term can be added to $\mat{A}_0$. Subtracting \refEq{variation-in-forward-ge4} from \refEq{adjoint-variation-R} gives
\begin{equation}\label{continuous-dR-minus-phidG}\begin{split}
\delta R &= \underbrace{\intxt{ \normp{\dudt{\vects{\phi}^{T}}\mat{A}_0 +\dudx{\vects{\phi}^{T}}\mat{A}_{1} +\vects{\phi}^{T}\mat{A}_2 -\vect{Q}^{T}}\delta\vect{W}}}_{\mathrm{P_1}} +\underbrace{\intxt{\vects{\phi}^T\frac{\partial \vect{S}}{\partial \vects{\omega}}\delta\vects{\omega}}}_{\mathrm{P_2}}\\
&+\underbrace{\intx{\vects{\phi}^{T}\mat{A}_0\delta\vect{W}|^{t_0} -\vects{\phi}^{T}\mat{A}_0\delta\vect{W}|^{t_1}}}_{\mathrm{P_3}} +\underbrace{\intt{\vects{\phi}^{T}\mat{A}_1\delta\vect{W}|^{x_0}- \vects{\phi}^{T}\mat{A}_1\delta\vect{W}|^{x_1}}  }_{\mathrm{P_4}}
\end{split}\end{equation}

The continuous adjoint equation is obtained with \refEq{continuous-dR-minus-phidG} by removing the dependency of $\mathrm{P}_1$ part on unknown $\delta \vect{W}$, i.e.
\begin{equation}\label{adjoint:transient-pde}
\mat{A}_{0}^{T}\dudt{\vects{\phi}} + \mat{A}_{1}^{T}\dudx{\vects{\phi}}  + \mat{A}_2^{T}\vects{\phi} =\vect{Q}
\end{equation}
The ICs for \refEq{adjoint:transient-pde} is obtained by by removing the dependency of $\mathrm{P}_3$ part on unknown $\delta \vect{W}$, i.e.
\begin{equation}\label{adjoint:transient-ics}
\vects{\phi}(x, t_1) = \vect{0}
\end{equation}
The BCs for \refEq{adjoint:transient-pde} are obtained by removing the dependency of $\mathrm{P}_4$ part on unknown $\delta \vect{W}$, i.e.
\begin{subequations}\label{adjoint:transient-bcs}\begin{align}
	B_2\normp{\vects{\phi}}=0, &\text{ for } x=x_0 \\
	B_1\normp{\vects{\phi}}, B_3\normp{\vects{\phi}}, B_4\normp{\vects{\phi}}, B_5\normp{\vects{\phi}}, B_6\normp{\vects{\phi}}=0, &\text{ for } x=x_1
\end{align}\end{subequations}
where $B_1$ to $B_6$ are functions of $\vects{\phi}$ that results from $\vects{\phi}^{T}\mat{A}_1$. Details of $B_1$ to $B_6$ are given in the Appendix B. Complete details for obtaining \refEq{adjoint:transient-pde}, \refEq{adjoint:transient-ics}, and \refEq{adjoint:transient-bcs} are referred to \cite{Hu2018PhdThesis}. 

Substituting \refEq{adjoint:transient-pde}, \refEq{adjoint:transient-ics}, and \refEq{adjoint:transient-bcs} into \refEq{continuous-dR-minus-phidG}, the variation in the response function is reduced to
\begin{equation}\label{continuous-tr-adjoint-sensitivity}
\delta R =\intxt{\vects{\phi}^T\frac{\partial \vect{S}}{\partial \vects{\omega}}\delta\vects{\omega}} + \intt{\normp{B_1\delta\alpha_g + B_3\delta T_l + B_4\delta T_g + B_5\delta u_l + B_6\delta u_g}|^{x_0} -\normp{B_2\delta p }|^{x_1}} + \intx{\vects{\phi}^{T}\mat{A}_0\delta\vect{W}|^{t_0}}
\end{equation}
Sensitivities of the response function to source parameters, boundary conditions, and initial conditions can be obtained with the first, second, and third part of the right hand side of \refEq{continuous-tr-adjoint-sensitivity}, respectively. In the following analysis, the sensitivity obtained with the continuous adjoint method is denoted as `CAS'. 
 
The same mesh, \refFig{schematic-mesh}, will be used to solve the adjoint equation. A central finite difference discretization will be used, which works well for problems where the forward solution is smooth, see \cite{Hu2018CAdjoint}. \refEq{adjoint:transient-pde} is discretized as
\begin{equation}\label{tr-continuous-adjoint-discretization-system}
\mat{A}_{0, i}^{T, n}\frac{\vects{\phi}_{i}^{n+1} - \vects{\phi}_{i}^{n}}{\Delta t} + \mat{A}_{1, i}^{T, n}\frac{\vects{\phi}_{i+1}^{n} - \vects{\phi}_{i-1}^{n}}{2\Delta x} + \mat{A}_{2, i}^{T, n}\vects{\phi}_{i}^{n} = \vect{Q}_i^{m}
\end{equation}
where the superscript $n$ denotes the $n^{\mathrm{th}}$ time step. Note that \refEq{adjoint:transient-ics} shows that the adjoint equation has to start from the end time ($t_1$) and integrate back to the start time $t_0$, thus \refEq{tr-continuous-adjoint-discretization-system} will be assembled to a global linear equation for solving $\vects{\phi}$ at step $n$. 

\subsection{Application of discrete adjoint method to two-fluid model}
The derivation of discrete adjoint equation starts with the discrete form of the forward problem, i.e. \refEq{backward-euler-fully-implicit}. For the $n^{\mathrm{th}}$ time step, The discrete form of \refEq{forward-governing-equation} can be summarized as
\begin{equation}\label{forward-governing-eq-tr-discrete}
\vect{G}_i^{n} = \vect{U}_i^{n}-\vect{U}_i^{n-1} + \vect{P}_{it,i}^{n}\normp{\alpha_{g,i}^{n}-\alpha_{g,i}^{n-1}} -\Delta t \opt{L}\normp{\vect{U}_i^{n}}=\vect{0}, n=1, 2, \cdots, \mathrm{M}
\end{equation}
where $ \mathrm{M}$ is the total number of time steps. Considering  \refEq{forward-governing-eq-tr-discrete} in all cells and the input parameters ($\vects{\omega}$), the discretized forward equation at $n^{\mathrm{th}}$ time step can be generalized as
\begin{equation}
\vect{G}^{n}(\vect{W}^{n}, \vect{W}^{n-1},\vects{\omega}) = \vect{0}, \text{ for } n=1, 2, \cdots, \mathrm{M}
\end{equation}
where $\vect{G}^{n}$ and $\vect{W}^{n}$ is the global residual and solution vector at $n^{\mathrm{th}}$ step, respectively. There are in total $\mathrm{M}$ steps. Let $\vect{G}$ be the global residual vector assembled with $\vect{G}^{n}$ for all $\mathrm{M}$ steps, i.e.
\begin{equation}
\vect{G} = \normp{\vect{G}^{1}, \vect{G}^{2}, \cdots, \vect{G}^{\mathrm{M}}}
\end{equation}
This way, the discrete form of the forward problem is specified by
\begin{equation}\label{transient-discrete-residual} 
\vect{G}\normp{\vect{W}, \vects{\omega}} = \vect{0}
\end{equation}
where $\vect{W}$ represents the vector of forward solution at all time steps and all cells. 

The response function for the discrete adjoint problem can be generalized as
\begin{equation}\label{transient-discrete-response-function}
R = R\normp{\vect{W}}
\end{equation}
where $R(\vect{W})$ is a scalar function of the solution vector $\vect{W}$. 

Let $\vects{\phi}^{n}$ be the adjoint solution at the $n^{\mathrm{th}}$ step and $\vects{\phi}$ be the global adjoint solution assembled from all time step, i.e.
\begin{equation}
\vects{\phi} = \normp{\vects{\phi}^{1}, \vects{\phi}^{2}, \cdots, \vects{\phi}^{\mathrm{M}}}
\end{equation}

The discrete adjoint equation is obtained by assigning $\vect{G}$ to the operator $\opt{G}$ and $R\normp{\vect{W}}$ to the operator $\opts{R}$ in \refEq{general-adjoint-equation}, i.e.
\begin{equation}\label{transient-discrete-adjoint-equation}
\normp{\frac{\partial \vect{G}}{\partial \vect{W}}}^{T}\vects{\phi} - \normp{\frac{\partial R}{\partial \vect{W}}}^{T} = \vect{0}
\end{equation}
and the sensitivity is given as
\begin{equation}\label{transient-discrete-adjoint-sensitivity}
\frac{\mathrm{d}R}{\mathrm{d}\vects{\omega}}= - \vects{\phi}^{T}\frac{\partial \vect{G}}{\partial \vects{\omega}}
\end{equation}
Because the effect of ICs and BCs has already been included in the residual vector $\vect{G}$, \refEq{transient-discrete-adjoint-equation} and \refEq{transient-discrete-adjoint-sensitivity} haven o explicit dependency on ICs and BCs. In the following analysis, the sensitivity obtained with the discrete adjoint method is denoted as `DAS'. 

The number of time steps for a transient problem may be very large, the dimension of \refEq{transient-discrete-adjoint-equation} is too large to solve efficiently . Fortunately, because the temporal discretization of the forward problem involves only forward solution at two steps, \refEq{transient-discrete-adjoint-equation} can be split into a series of linear equations with a smaller dimension. It is found that the coefficient matrix/vector in \refEq{transient-discrete-adjoint-equation} has the following form
\begin{equation}
\normp{\frac{\partial \vect{G}}{\partial \vect{W}}}^{T} = \begin{pmatrix}
\normp{\frac{\partial \vect{G}^{1}}{\partial \vect{W}^{1}}}^{T} & \normp{\frac{\partial \vect{G}^{2}}{\partial \vect{W}^{1}}}^{T} & 0& 0  & \cdots & 0 \\
0 & \normp{\frac{\partial \vect{G}^{2}}{\partial \vect{W}^{2}}}^{T} & \normp{\frac{\partial \vect{G}^{3}}{\partial \vect{W}^{2}}}^{T} & 0 & \cdots & 0 \\
\vdots & \vdots & \vdots & \vdots & \vdots & \vdots \\
0 & \cdots & 0 & 0 &\normp{\frac{\partial \vect{G}^{\mathrm{M}-1}}{\partial \vect{W}^{\mathrm{M}-1}}}^{T} & \normp{\frac{\partial \vect{G}^{\mathrm{M}}}{\partial \vect{W}^{\mathrm{M}-1}}}^{T} \\
0 & \cdots & 0 & 0 & 0 &\normp{\frac{\partial \vect{G}^{\mathrm{M}}}{\partial \vect{W}^{\mathrm{M}}}}^{T}
\end{pmatrix},\normp{\frac{\partial R}{\partial \vect{W}}}^{T} = \begin{pmatrix}
\normp{\frac{\partial R}{\partial \vect{W}^{1}}}^{T}\\
\normp{\frac{\partial R}{\partial \vect{W}^{2}}}^{T}\\
\vdots\\
\normp{\frac{\partial R}{\partial \vect{W}^{\mathrm{M}-1}}}^{T}\\
\normp{\frac{\partial R}{\partial \vect{W}^{\mathrm{M}}}}^{T}\\
\end{pmatrix}
\end{equation}

Thus, the adjoint variable can be solved with
\begin{equation}\label{transient-discrete-adjoint-solution-method}\begin{split}
\normp{\frac{\partial \vect{G}^{\mathrm{M}}}{\partial \vect{W}^{\mathrm{M}}}}^{T} \vects{\phi}^{\mathrm{M}} &= \normp{\frac{\partial R}{\partial \vect{W}^{\mathrm{M}}}}^{T} \\
\normp{\frac{\partial \vect{G}^{n}}{\partial \vect{W}^{n}}}^{T} \vects{\phi}^{n} &= \normp{\frac{\partial R}{\partial \vect{W}^{n}}}^{T} - \normp{\frac{\partial \vect{G}^{n+1}}{\partial \vect{W}^{n}}}^{T} \vects{\phi}^{n+1}, \text{ for } n = \mathrm{M}-1 , \cdots, 1
\end{split}\end{equation}
In practice, the coefficient matrices and vectors in \refEq{transient-discrete-adjoint-solution-method} are obtained with a numerical differentiation scheme. After obtaining the adjoint variable, the sensitivity can be obtained with
\begin{equation}\label{transient-discrete-adjoint-sensitivity-reduced}
\frac{\mathrm{d}R}{\mathrm{d}\vects{\omega}}=  -\sum_{n=1}^{\mathrm{M}} \normp{\vects{\phi}^{n}}^{T}\frac{\partial \vect{G}^{n}}{\partial \vects{\omega}}
\end{equation}

Because the forward discretization is upwind and the coefficient matrices for the DAS is the transpose of the forward Jacobian matrices, the solver for solving the discrete adjoint equation behaves like the forward solver. In other words, if the forward solver is stable and robust, the discrete adjoint solver is also stable and robust. This is a critical feature of the discrete adjoint method. 
\subsection{Steady-state problems}
The adjoint equations for steady-state problems can be generalized from \refEq{adjoint:transient-pde} and \refEq{transient-discrete-adjoint-equation} by removing the dependency on time. 

\subsection{Response function}
The form of the response function does not affect the application of the adjoint method and should depend on different problems. This article considers a group of response function that can be written as
\begin{equation}
R(x_d, t_d) =\left\{
\begin{tabular}{ll}
 $\intxt{\xi(x; x_d, t_d)q(x)}$, & continuous form \\
$\sum_{i=1}^{\mathrm{N}}\sum_{n=1}^{\mathrm{M}} \xi(x_i; x_d, t_d)q_i \Delta x \Delta t$, & dicrete form
\end{tabular}\right.\end{equation}
where $\mathrm{N}$ is the total number of cells, $\mathrm{M}$ is the total number of time steps, $x_i$ is the location of the $i^{\mathrm{th}}$ cell, and $q$ denotes a function of the primitive variable, e.g. $q = \alpha_g$. The other variable, $\xi(x)$, is a weight function which is used to study the behavior of the response function at different location $x_d$ and time $t_d$. The dependency of the response function on $x_d$  and $t_d$ is designed for verification purposes. In this study, $\xi(x)$ is non-zero only in the neighboring cells of $x_d$ and at time step $t_d$, which simulates a point-wise response function. The adjoint sensitivity analysis is in general efficient for problems with few response functions. 

The sensitivity coefficient is used for comparison. The sensitivity coefficient is defined as
\begin{equation}
\mathrm{SC} = \frac{\mathrm{d}R}{\mathrm{d}\omega}\frac{\omega_{0}}{R_{0}}
\end{equation}
where $\omega_{0}$ and $R_{0}$ are the nominal values of the input parameter and response, respectively. 

\subsection{Perturbation sensitivity analysis}
It is in general impractical to get exact sensitivities to verify the sensitivities obtained with the adjoint method. A perturbation analysis is used to provide the reference value of the sensitivities. The procedure for the perturbation sensitivity analysis is straightforward, 
\begin{equation}
\mathrm{PS} \approx \frac{R_{\varepsilon} - R_0}{\varepsilon}
\end{equation}
where $\mathrm{PS}$ denotes the sensitivity from perturbation sensitivity analysis, $R_{\varepsilon}$ and $R_{0}$ are the response functions calculated by performing the forward simulation with and without the perturbation to the input parameter of interest, respectively. In practice, $\varepsilon = 10^{-6}$ is used. 

\section{Numerical tests}\label{sec-four}
\subsection{Steady-state test with faucet flow}\label{sec-four-p1}
This test is the Ransom's faucet flow problem \cite{dinh2003understanding, zou2015, hewitt1986, hewitt2013}, which has an analytical solution. This test problem consists of a liquid stream entering a vertical tube at the top and falling under gravity to form a liquid stream of decreasing cross-section. The length of the vertical tube is $L = 12$ m.  The boundary conditions are: $\alpha_{g, inlet} = 0.2$, $T_{l, inlet} = 300$ K, $T_{g, inlet} = 500$ K, $u_{l, inlet} = 10$ m/s, $u_{g, inlet} = 0$ m/s, and $p_{outlet} = 0.1$ MPa. Properties of liquid and gas are obtained from the IAPWS-IF97 formulation \cite{iapws1998}. Since there is no mass and heat transfer between the liquid and gas phases, the superheated steam is used to simulate the gas phase. The source vector of this problem is
\begin{equation}
\vect{S}=\begin{pmatrix}
0 & \alpha_l \rho_l g & 0  & 0 & \alpha_g \rho_g g & 0 \\
\end{pmatrix}^{T}
\end{equation}
where $g$ is the gravitational acceleration constant and the superscript $T$ is the transpose operator.   

The void fraction, liquid velocity, and pressure are of particular interest to this test. The steady-state solution \cite{zou2015, zou2016new} of this problem is
\begin{align}
u_{l}(x)   &= \sqrt{u_{l, inlet}^2 + 2 g_{e}x} \\
\alpha_{g}(x) &= 1.0 - \frac{\normp{1-\alpha_{g, inlet}}u_{l, inlet}}{u_{l, ss}(x)} \\
p(x)      &= p_{outlet} - \rho_g g(L - x)
\end{align}
where $g$ is the gravitational acceleration constant and $g_{e}$ is
\begin{equation}
g_{e} = g\normp{1 - \rho_g/\rho_l}
\end{equation}
where $\rho_g = 0.435$ $\mathrm{kg/m^3}$ and $\rho_l = 996.56$ $\mathrm{kg/m^3}$.

\subsubsection{Input parameters}
For this test, 3 input parameters are considered: inlet void fraction, inlet liquid velocity, and gravitational constant, i.e.
\begin{equation}
\vects{\omega} = \begin{pmatrix}
\alpha_{g,inlet} & u_{l, inlet} & g \\
\end{pmatrix}
\end{equation} 
The inlet void fraction and inlet liquid velocity represent typical input parameters related to boundary conditions; while the gravitational constant represent typical input parameters in the source terms. Several sensitivities can be obtained analytically, i.e.
\begin{equation}\label{SS-Faucet-Analytical-Sensitivity}\begin{split}
\frac{\mathrm{d}\alpha_{g}(x)}{\mathrm{d}\alpha_{g, inlet}} &= \frac{u_{l, inlet}}{u_{l}(x)}, \frac{\mathrm{d}\alpha_{g}(x)}{\mathrm{d}u_{l, inlet}} = -\frac{2\normp{1-\alpha_{g, inlet}}g_e x}{u_{l}^3(x)}, \frac{\mathrm{d}\alpha_{g}(x)}{\mathrm{d}g} = \frac{\normp{1-\alpha_{g, inlet}}u_{l, inlet} x}{u_{l}^3(x)} \\
\frac{\mathrm{d}u_{l}(x)}{\mathrm{d}u_{l, inlet}} &= \frac{u_{l, inlet}}{u_{l}(x)}, \frac{\mathrm{d}u_{l}(x)}{\mathrm{d}g} = \frac{\normp{1 - \rho_g/\rho_l}x}{u_{l}(x)},\frac{\mathrm{d}p(x)}{\mathrm{d}g} = - \rho_g\normp{L-x}
\end{split}\end{equation}
These analytical sensitivities are used to verify the adjoint sensitivity analysis framework by studying the estimation error in the continuous and discrete adjoint sensitivity analysis. 

\subsubsection{Results}
\begin{figure}[!htbp]
	\centering
	\begin{subfigure}[t]{0.45\textwidth}
		\centering
		\includegraphics[width=\textwidth, height=0.8\textwidth]{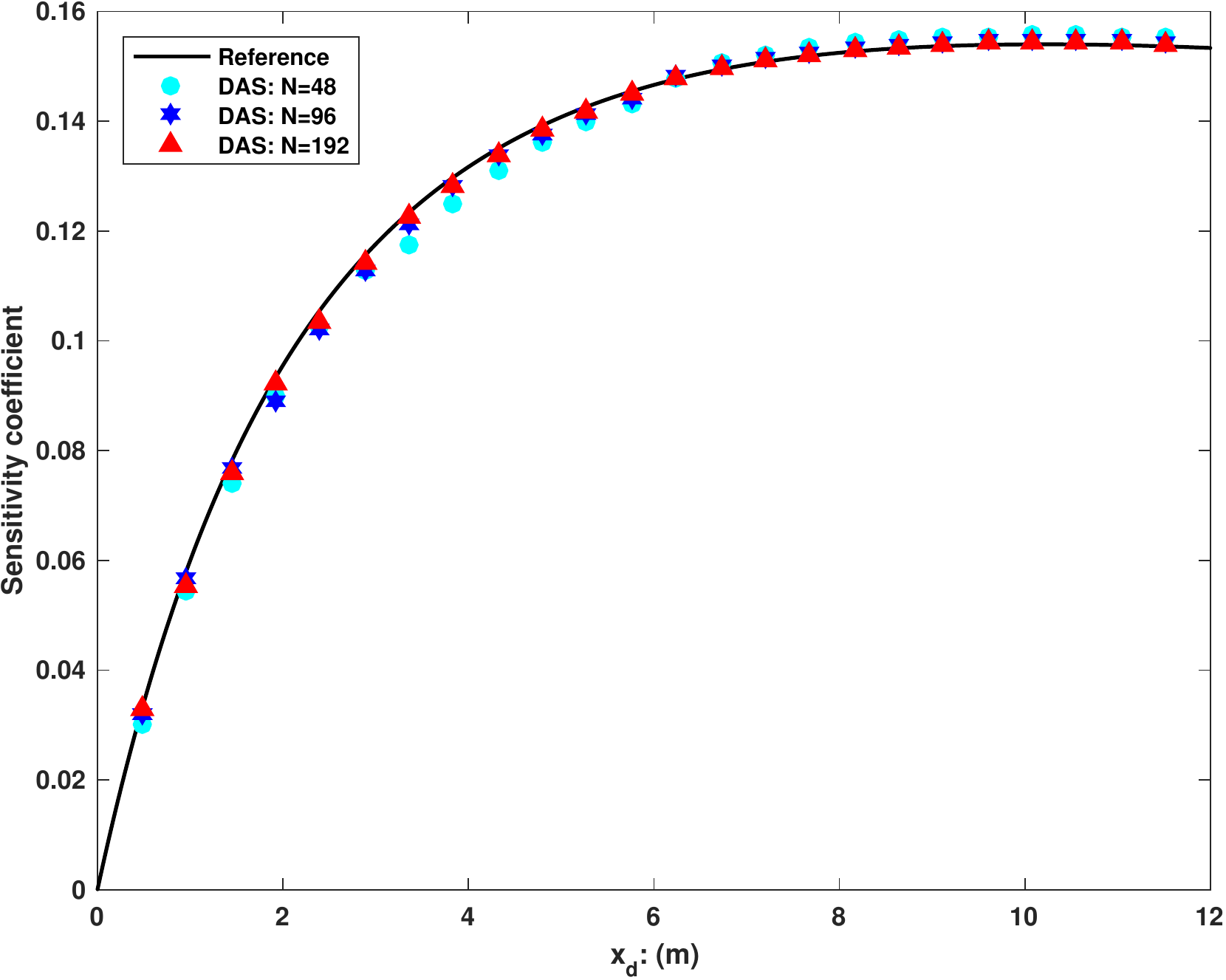}
		\caption{DAS: $q=\alpha_g, \omega = g$}
	\end{subfigure}%
	~
	\begin{subfigure}[t]{0.45\textwidth}
		\centering
		\includegraphics[width=\textwidth, height=0.8\textwidth]{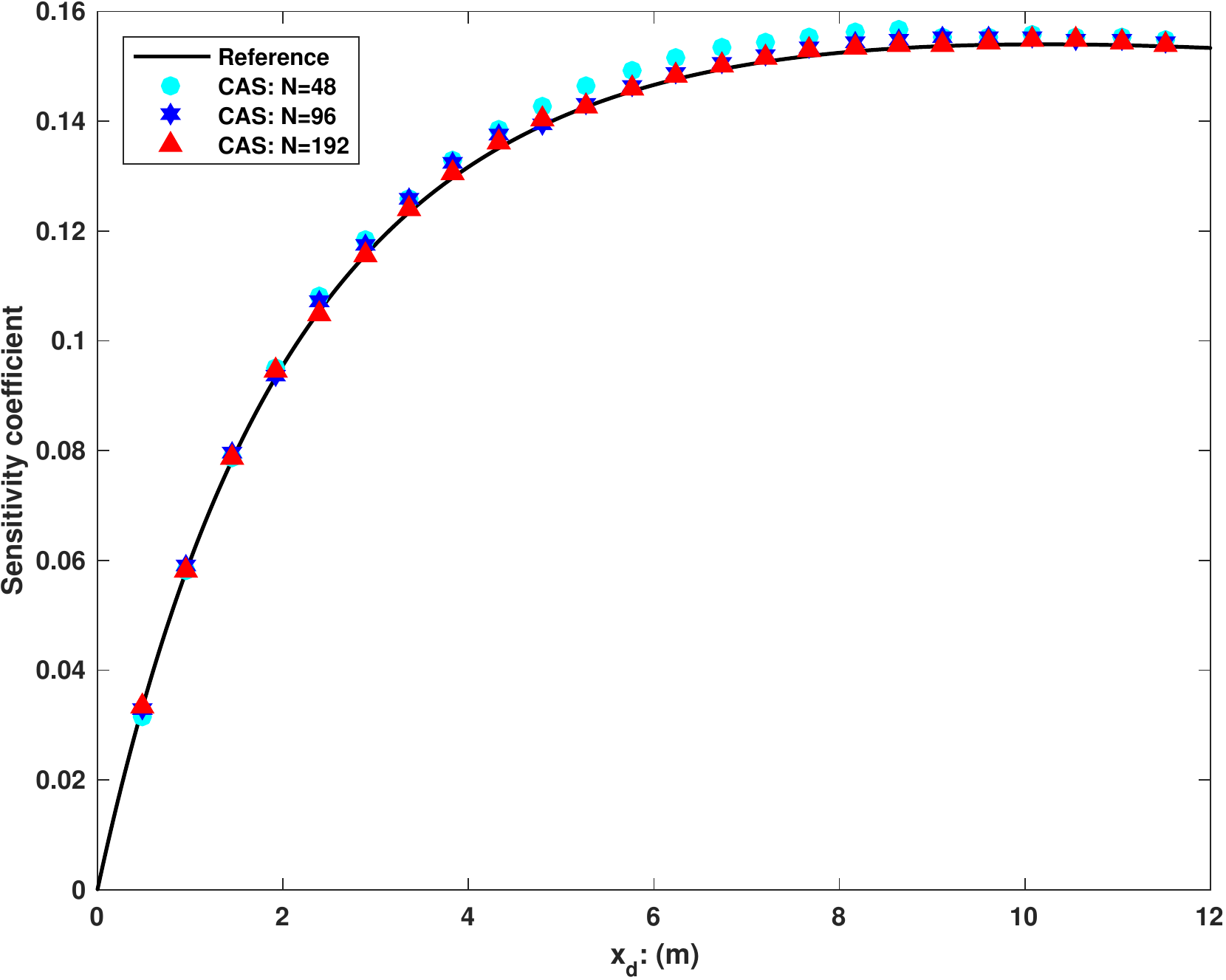}
		\caption{CAS: $q=\alpha_g, \omega = g$}
	\end{subfigure}%
	\caption{Mesh convergence of continuous and discrete adjoint method for faucet flow at steady-state} \label{SS-Faucet-SA-VoidF-Conv}
\end{figure}

The forward solver is at first run to reach steady-state for preparing the coefficient matrices/vectors. Numerical solutions match the analytical solution well. Assessment of the forward solver is omitted in this article and is referred to \cite{Hu2018Implicit}. For this test, $\sigma_d = 0.25$ and 11 points ($x_d$) are used. At first, a mesh convergence study is done to select an appropriate mesh size, shown in \refFig{SS-Faucet-SA-VoidF-Conv}. It is seen that $\mathrm{N} =192$ is fine enough for the sensitivity analysis, which is used in the following analysis. Then both continuous and discrete adjoint method are applied to calculate the sensitivities given in \refEq{SS-Faucet-Analytical-Sensitivity}. The estimation error for the sensitivities from both methods is listed in \refTab{SS-Faucet-SA-Error}. It is seen that both continuous and discrete adjoint methods calculate the sensitivities very well. From the comparison, it is seen that the continuous adjoint method tends to give more accurate sensitivities than the discrete adjoint method. The explanation is that the coefficient matrix is approximated with a finite difference method in the discrete adjoint method; while the coefficient matrix is given analytically in the continuous adjoint method. Because of the extra effort to compute the coefficient matrix with a finite difference method, the discrete adjoint method takes more computational time than the continuous adjoint method. 

\begin{table}[!htbp]
	\centering
	\caption{Error analysis of the sensitivity coefficient from continuous and discrete adjoint method for faucet flow at steady-state. $E_{\mathrm{DAS}}$ and $E_{\mathrm{CAS}}$ represent the relative error in the sensitivities from DAS and CAS. The analytical sensitivities are used as the reference values. }\label{SS-Faucet-SA-Error}
	\begin{tabular}{c|ccc|ccc|ccc}
		\hline
		& REF & $E_{\mathrm{DAS}}$: \% & $E_{\mathrm{CAS}}$: \% & REF & $E_{\mathrm{DAS}}$: \% & $E_{\mathrm{CAS}}$: \% & REF & $E_{\mathrm{DAS}}$ & $E_{\mathrm{CAS}}$: \% \\ \hline
		$x_d$:(m) &  \multicolumn{3}{|c|}{$q = \alpha_g, \omega = \alpha_{g, inlet}$} &\multicolumn{3}{|c|}{$q = \alpha_g, \omega = u_{l, inlet}$}&\multicolumn{3}{|c}{$q = \alpha_g, \omega = g$} \\ \hline
		0.96	&	9.17E-01	&	0.94	&	-0.03	&	-1.16E-01	&	-5.47	&	-10.48	&	5.82E-02	&	-4.60	&	-0.40	\\
		1.92	&	8.52E-01	&	0.89	&	-0.24	&	-1.87E-01	&	-1.99	&	-4.05	&	9.33E-02	&	-1.25	&	1.55	\\
		2.88	&	7.99E-01	&	1.12	&	0.68	&	-2.31E-01	&	-1.92	&	-0.97	&	1.15E-01	&	-1.28	&	0.05	\\
		3.84	&	7.55E-01	&	1.25	&	0.17	&	-2.60E-01	&	-1.66	&	-2.62	&	1.30E-01	&	-1.09	&	0.55	\\
		4.80	&	7.18E-01	&	1.02	&	-0.01	&	-2.78E-01	&	-1.02	&	-1.80	&	1.39E-01	&	-0.50	&	0.82	\\
		5.76	&	6.85E-01	&	1.10	&	0.56	&	-2.91E-01	&	-0.91	&	-0.64	&	1.45E-01	&	-0.43	&	0.28	\\
		6.72	&	6.57E-01	&	1.02	&	0.36	&	-2.99E-01	&	-0.69	&	-0.48	&	1.49E-01	&	0.20	&	0.37	\\
		7.68	&	6.32E-01	&	1.04	&	0.11	&	-3.04E-01	&	-0.61	&	-1.14	&	1.52E-01	&	0.19	&	0.59	\\
		8.64	&	6.09E-01	&	1.06	&	0.50	&	-3.06E-01	&	-0.52	&	-0.48	&	1.53E-01	&	0.20	&	0.26	\\
		9.60	&	5.89E-01	&	0.86	&	0.34	&	-3.08E-01	&	-0.42	&	-0.44	&	1.54E-01	&	0.23	&	0.26	\\
		10.56	&	5.71E-01	&	0.89	&	0.18	&	-3.08E-01	&	-0.37	&	-0.84	&	1.54E-01	&	0.23	&	0.45	\\
		11.52	&	5.54E-01	&	0.92	&	0.46	&	-3.07E-01	&	-0.31	&	-0.41	&	1.54E-01	&	0.25	&	0.23	\\
		\hline
		$x_d$:(m) &  \multicolumn{3}{|c|}{$q = u_l, \omega = u_{l, inlet}$} &\multicolumn{3}{|c|}{$q = u_l, \omega = g$}&\multicolumn{3}{|c}{$q = p, \omega = g$} \\ \hline
		0.96	&	9.17E-01	&	0.73	&	0.82	&	8.64E-02	&	-5.04	&	-3.30	&	-4.71E-04	&	1.96	&	-0.89	\\
		1.92	&	8.52E-01	&	0.68	&	0.63	&	1.61E-01	&	-1.29	&	0.26	&	-4.30E-04	&	1.67	&	-0.83	\\
		2.88	&	7.99E-01	&	0.96	&	0.30	&	2.26E-01	&	-1.71	&	0.31	&	-3.89E-04	&	1.97	&	0.13	\\
		3.84	&	7.55E-01	&	1.14	&	1.05	&	2.85E-01	&	-1.80	&	-0.78	&	-3.48E-04	&	2.32	&	-0.11	\\
		4.80	&	7.18E-01	&	0.96	&	0.88	&	3.38E-01	&	-0.81	&	0.04	&	-3.07E-04	&	1.88	&	-0.60	\\
		5.76	&	6.85E-01	&	1.09	&	0.27	&	3.87E-01	&	-0.97	&	0.14	&	-2.66E-04	&	2.21	&	-0.01	\\
		6.72	&	6.57E-01	&	0.95	&	0.06	&	4.33E-01	&	-0.19	&	0.56	&	-2.25E-04	&	1.49	&	-0.80	\\
		7.68	&	6.32E-01	&	1.03	&	1.00	&	4.76E-01	&	-0.38	&	0.01	&	-1.84E-04	&	1.89	&	-0.71	\\
		8.64	&	6.09E-01	&	1.10	&	0.19	&	5.16E-01	&	-0.51	&	0.09	&	-1.43E-04	&	9.67	&	0.07	\\
		9.60	&	5.89E-01	&	0.95	&	0.03	&	5.55E-01	&	-0.18	&	0.37	&	-1.02E-04	&	0.96	&	-1.49	\\
		10.56	&	5.71E-01	&	1.02	&	1.07	&	5.91E-01	&	-0.30	&	-0.01	&	-6.14E-05	&	1.84	&	-1.05	\\
		11.52	&	5.54E-01	&	1.08	&	0.13	&	6.26E-01	&	-0.41	&	0.06	&	-2.05E-05	&	6.25	&	3.86	\\
		\hline
	\end{tabular}
\end{table}

\begin{figure}[!htbp]
	\centering
	\begin{subfigure}[t]{0.45\textwidth}
		\centering
		\includegraphics[width=\textwidth, height=0.8\textwidth]{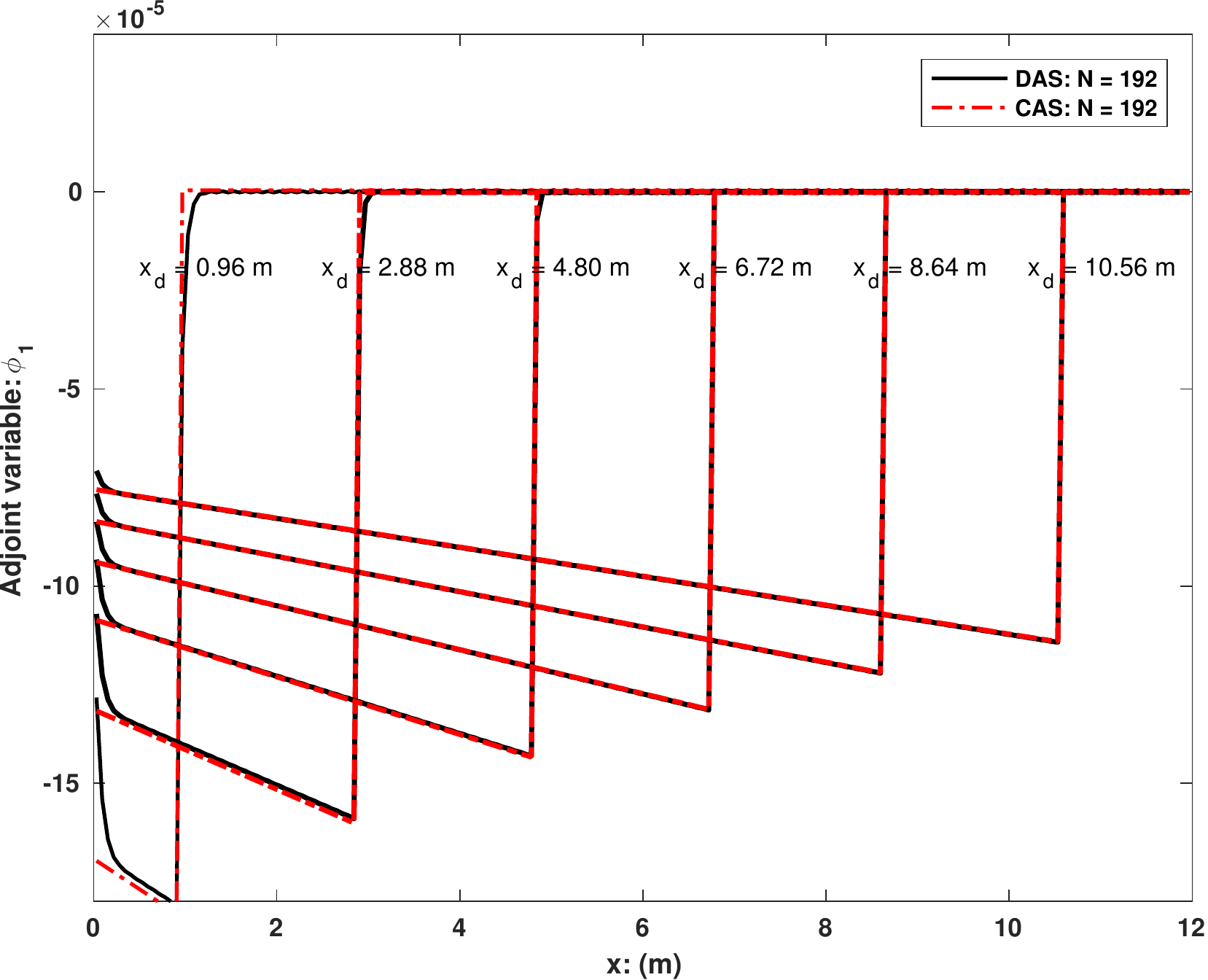}
		\caption{$\phi_1$ for $q=\alpha_g$}
	\end{subfigure}%
	~
	\begin{subfigure}[t]{0.45\textwidth}
		\centering
		\includegraphics[width=\textwidth, height=0.8\textwidth]{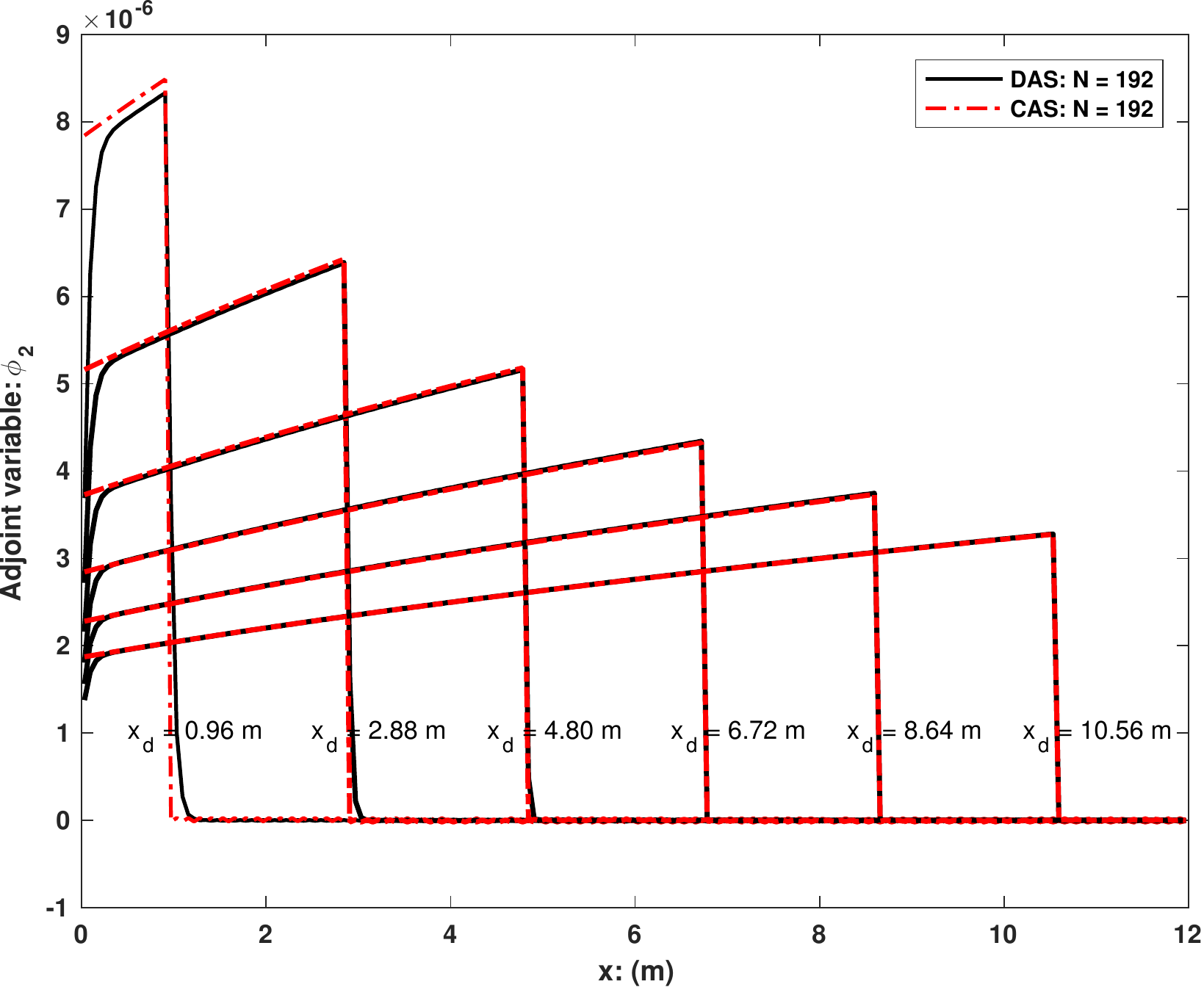}
		\caption{$\phi_2$ for $q=\alpha_g$}
	\end{subfigure}%
	
	\begin{subfigure}[t]{0.45\textwidth}
		\centering
		\includegraphics[width=\textwidth, height=0.8\textwidth]{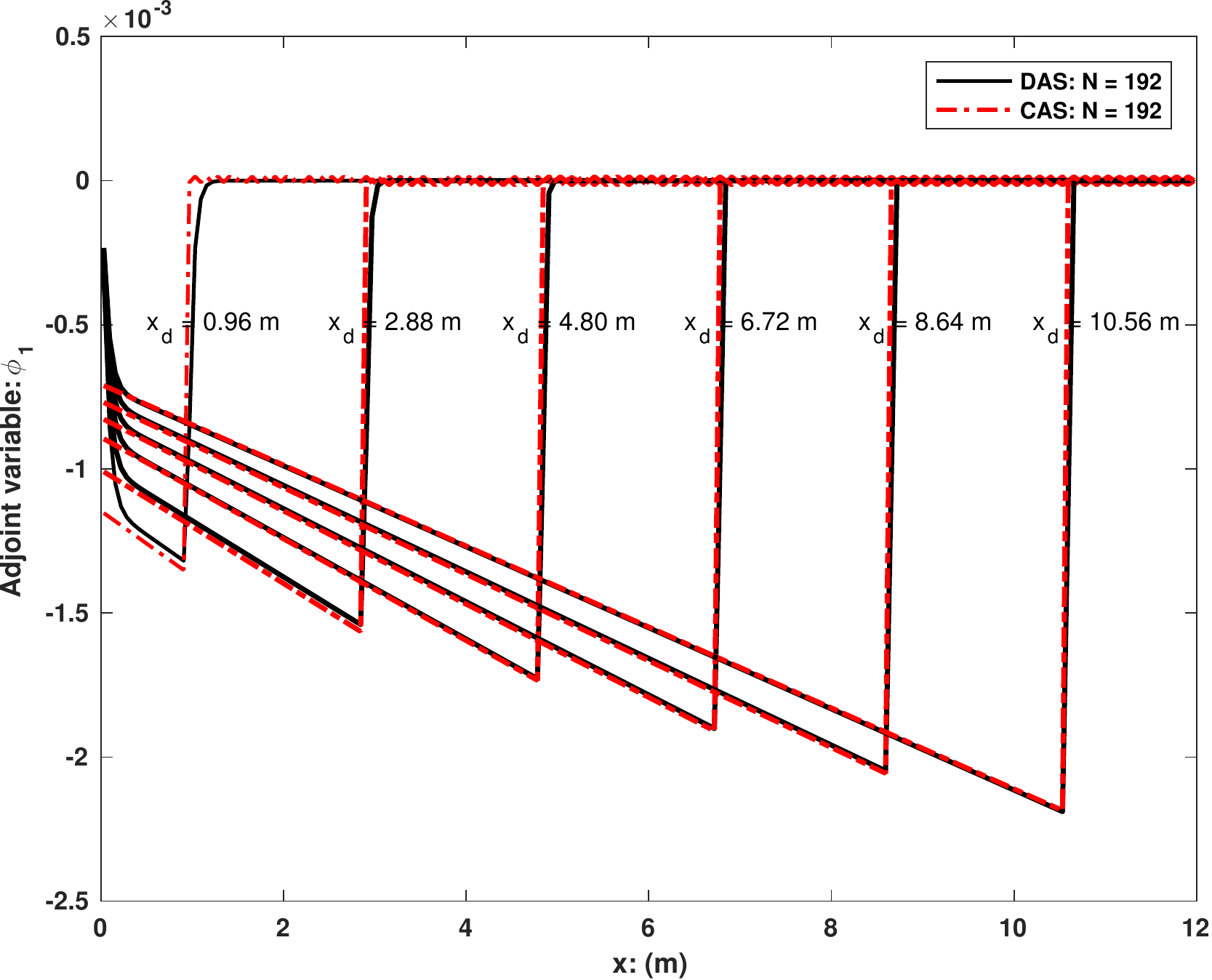}
		\caption{$\phi_1$ for $q=u_l$}
	\end{subfigure}%
	~
	\begin{subfigure}[t]{0.45\textwidth}
		\centering
		\includegraphics[width=\textwidth, height=0.8\textwidth]{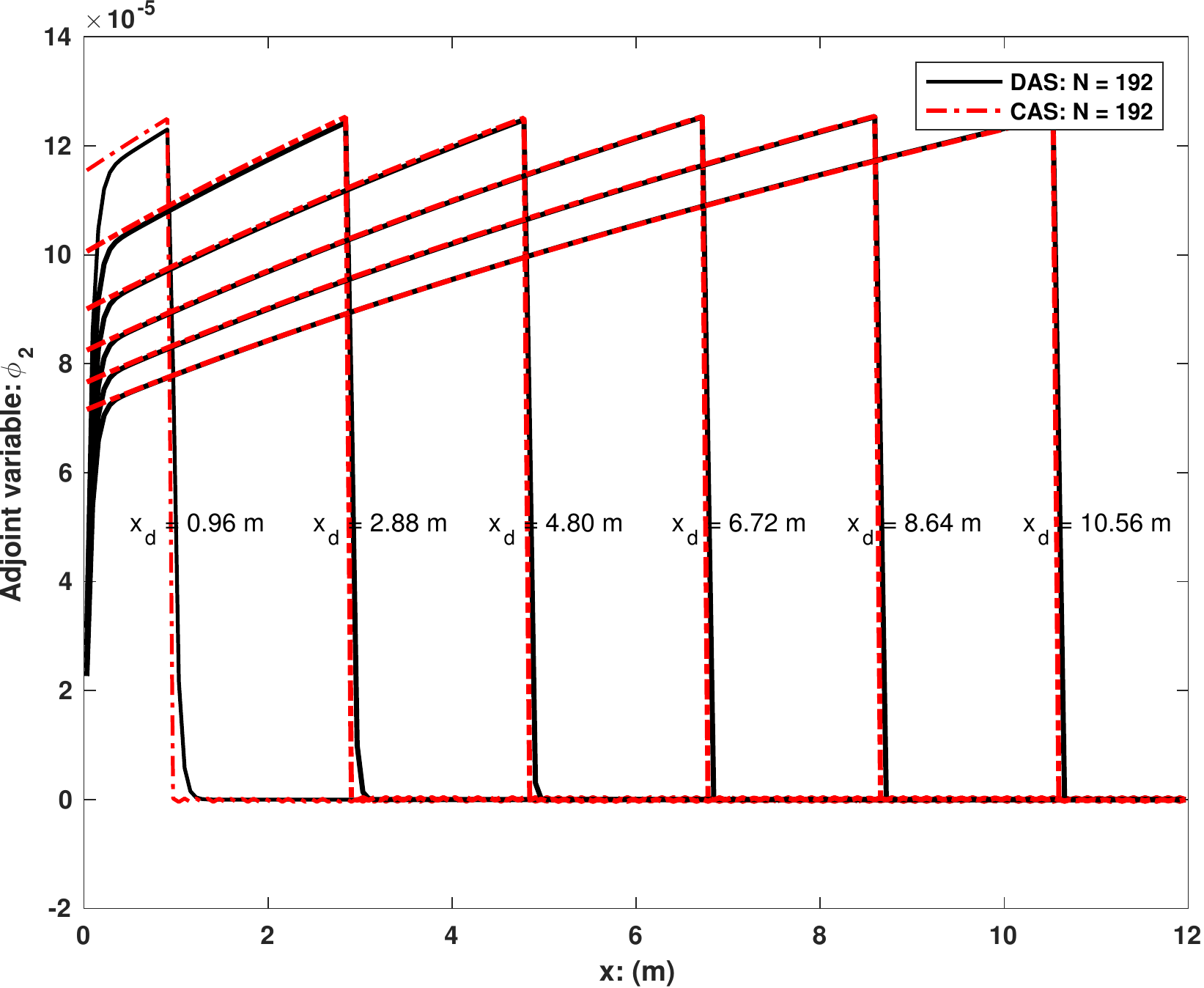}
		\caption{$\phi_2$ for $q=u_l$}
	\end{subfigure}%
	\caption{Adjoint variable from continuous and discrete adjoint method for faucet flow at steady-state } \label{SS-Faucet-SA-AdjointSolution}
\end{figure}

\refFig{SS-Faucet-SA-AdjointSolution} shows the adjoint variable components obtained from both continuous and discrete adjoint method. The adjoint variable component represents the effect (or value) of the respective forward governing equation to the response of interest, e.g. $\phi_1$ and $\phi_2$ represents the effect of the liquid mass and momentum equation, respectively. For this faucet flow problem, it is the liquid mass and momentum equation that affect the forward solution (and the response). The effect of the other 4 equations is negligible. The profile of the adjoint component is reasonable. For a response at a certain location ($x_d$), we expect that the response is mainly affected by its upstream but not its downstream. The nonzero and zero magnitude of the adjoint component at the upstream and downstream, respectively, is consistent with this expectation. It is seen that the adjoint variable obtained from continuous and discrete adjoint method matches well except near the boundaries. This is reasonable. For the continuous adjoint method, the effect of boundary conditions is calculated by \refEq{continuous-tr-adjoint-sensitivity} (i.e. the $B\normp{\vects{\phi}}$ variable), which depends only on the adjoint variable at the boundary. However, for the discrete adjoint method, the effect of boundary conditions is calculated by \refEq{transient-discrete-adjoint-sensitivity}, which actually depends on both the boundary point and its neighbors because of the three-point stencil used in the forward numerical scheme.

\subsection{Transient test with BFBT benchmark}\label{sec-four-p2}
One of the most valuable and publicly available databases for thermal-hydraulic modeling of Boiling Water Reactor (BWR) channels is the OECD/NEA BWR Full-size Fine-mesh Bundle Test (BFBT) benchmark, which includes sub-channel void fraction measurements in a full-scale BWR fuel assembly \cite{bfbtV1}. There are two types of void distribution measurement systems: an X-ray CT scanner and an X-ray densitometer \cite{bfbtV1}. There are 4 measurement locations, which are denoted by DEN \#3 (0.682 m), DEN \#2 (1.706 m), DEN \#1 (2.730 m), and CT (3.708 m) starting from the bottom. The geometry and system configurations of the channel are shown in \refTab{forward:TabBFBT-PC}.

\begin{table}[!htb]
	\caption{Experiment conditions for BFBT benchmark}\label{forward:TabBFBT-PC}
	\centering
	\begin{tabular}{ll|ll}
		\hline
		\multicolumn{2}{c|}{Geometry parameters} & \multicolumn{2}{c}{System/experiment parameters} \\ \hline
		Heated length (m) & 3.708 & Pressure (MPa) & 3.9 - 8.7 \\
		Width of channel box (m) & 0.1325 & Inlet temperature ($^{\circ}$C) & 238. - 292. \\
		Hydraulic diameter (m) & 0.01284  & Inlet subcooling (kJ/kg) & 50. - 56. \\
		Volumetric wall surface area ($\mathrm{m^{-1}}$) & 311.5 & Flow rate (t/h) & 10. - 70. \\
		Flow area ($\mathrm{m}^{2}$)& 9.463E-03  & Power (MW) & 0.62 - 7.3 \\
		Wetted perimeter (m) & 3.003 & Exit quality (\%) & 8. - 25. \\
		\hline
	\end{tabular}
\end{table}

Closure correlations are required to simulate the behavior of a boiling system. For this type of problems, the source vector $\vect{S}$ is modeled as
\begin{equation}
\vect{S} = \begin{pmatrix}
-\Gamma_g \\
-\alpha_l\rho_l g -f_{wl} + f_{i} - \Gamma_g u_{i} \\
Q_{wl} + Q_{il} - \Gamma_w h_{l}^{'}-\Gamma_{ig}h_{l}^{*} + \normp{f_{i}-f_{wl}-\alpha_l\rho_l g - \Gamma_g u_{i}}u_l + \Gamma_g\frac{u_l^2}{2}\\
\Gamma_g \\
-\alpha_g\rho_g g -f_{wg} - f_{i} + \Gamma_g u_{i} \\
Q_{wg} + Q_{ig} + \Gamma_w h_{g}^{'}+\Gamma_{ig}h_{g}^{*} + \normp{-f_{i}-f_{wg}-\alpha_g\rho_g g + \Gamma_g u_{i}}u_g - \Gamma_g\frac{u_g^2}{2}\\
\end{pmatrix}
\end{equation}
where $\Gamma_g$ is the net vapor generation rate due to near wall vapor generation ($\Gamma_w$) and bulk vapor generation ($\Gamma_{ig}$), $u_i$ is the interface velocity, $f_i$ is the interfacial friction, $f_{wk}$ is the phasic wall friction, $Q_{ik}$ is the phasic interfacial heat flux, $Q_{wk}$ is the phasic wall heat flux, $h_k^{'}$ is the phasic enthalpy carried by the wall vapor generation term ($\Gamma_{w}$), and $h_k^{*}$ is the phasic enthalpy carried by the bulk vapor generation term ($\Gamma_{ig}$). Correlations based on RELAP5-3D code manual \cite{RELAP5V1, RELAP5V4} are used to model these variables. Details of all physical models are provided in \cite{Hu2017Riemann, Hu2018PhdThesis}.

The void fraction is the main response of interest in this test. The main physical model that affects the void fraction is the near wall vapor generation rate ($\Gamma_w$), which is determined by the Lahey (\cite{lahey1978mechanistic}) model, i.e.
\begin{align}\label{Saha-Zuber-Lahey-model}
\Gamma_w &= \left\{ 
\begin{tabular}{ll}
$0$; & $h_l < h_{cr}$\\
$\frac{q_{w}a_{w}\normp{h_l - h_{cr}}}{\normp{h_{l, \mathrm{sat}}-h_{cr}}\normp{1+\varepsilon_p} h_{lg}}$;& $h_{cr}\le h_l \le h_{l, \mathrm{sat}}$\\
$\frac{q_{w}a_{w}}{h_{lg}}$;& $h_l > h_{cr}$
\end{tabular} \right. \\
\varepsilon_p &= \frac{\rho_l \normb{h_{l, \mathrm{sat}} - \min\normp{h_l, h_{l, \mathrm{sat}}}}}{\rho_g h_{lg}}
\end{align}
where $q_w$ is the volumetric wall heat flux, $a_w$ is the volumetric wall surface area, $h_{l, \mathrm{sat}}$ is the specific saturation enthalpy of liquid phase, and $h_{lg}$ is the difference between the liquid and gas phase specific saturation enthalpy. In case of the the subcooled boiling, the bulk liquid can be subcooled while liquid in the boundary layer is saturated and is flashing to vapor. Depending on the inlet subcooling level, the system may start from a liquid phase ($\alpha_g = 0$) and transits to two phases ($\alpha_g \neq 0$). The condition for this transition is that the specific enthalpy of the liquid phase ($h_l$) is greater than a critical enthalpy $h_{cr}$. Passing this transition point, the void fraction begins to increase. Currently, the critical enthalpy ($h_{cr}$) is modeled with the Saha-Zuber \cite{RELAP5V4} correlation, i.e.
\begin{equation}\label{Saha-Zuber-model}
h_{cr}=\left\{
\begin{tabular}{ll}
$h_{l, \mathrm{sat}} - \frac{C_{p,l}\cdot \mathrm{St} }{\mathrm{St}_{cr}}$; & $\mathrm{Pe}<70,000$\\
$ h_{l, \mathrm{sat}} - \frac{C_{p,l}\cdot \mathrm{Nu} }{\mathrm{Nu}_{cr}}$; & $\mathrm{Pe}\ge70,000$
\end{tabular}\right.
\end{equation}
where $\mathrm{Pe}$ is the Peclet number, $\mathrm{St}$ is the Stanton number, and $\mathrm{Nu}$ is the Nusselt number. More details are referred to \cite{RELAP5V4}. 

\subsection{Transient test problem}
\begin{figure}[!htbp]
	\centering
	\begin{subfigure}[t]{0.32\textwidth}
		\centering
		\includegraphics[width=\textwidth, height=0.8\textwidth]{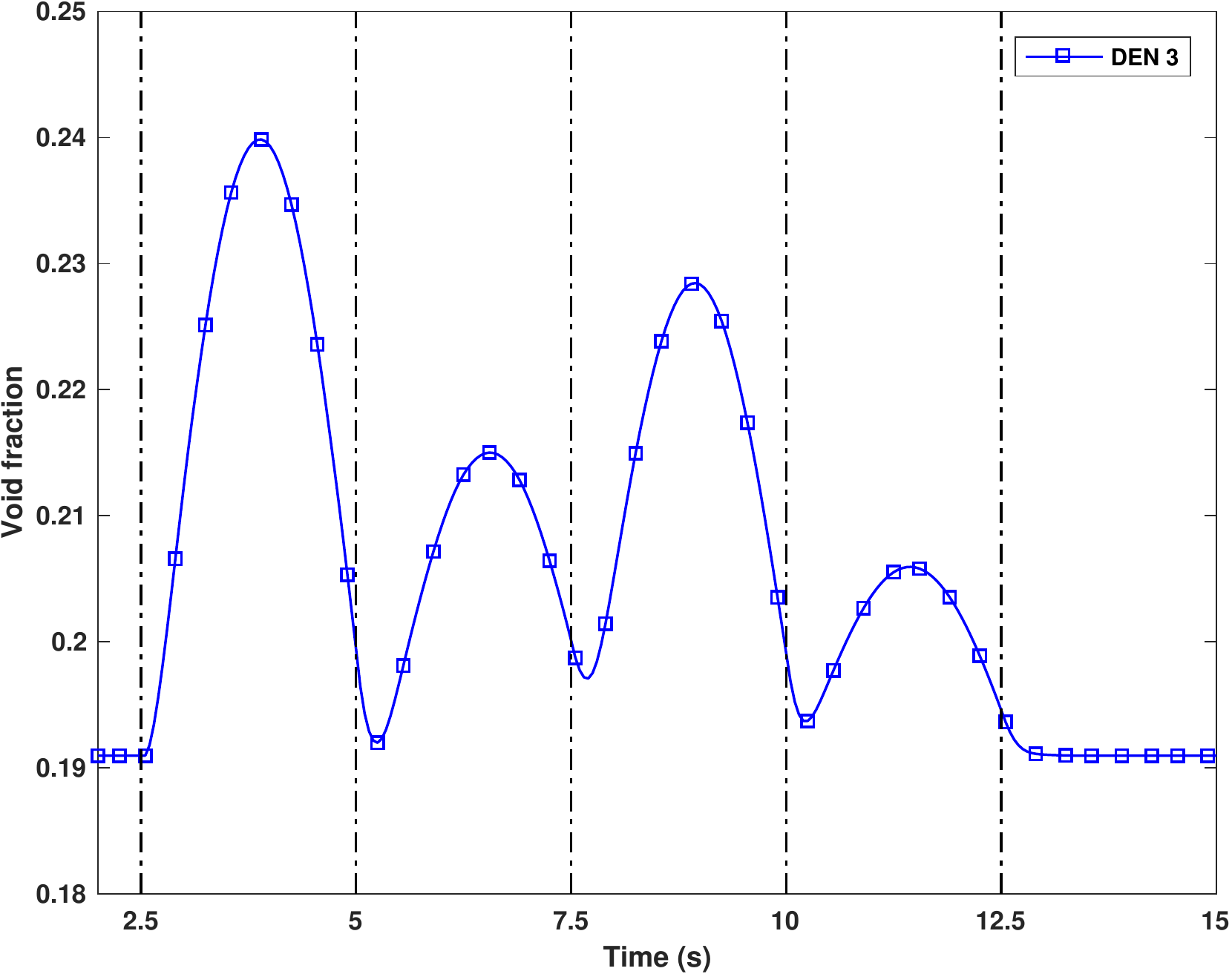}
		\caption{DEN 3}
	\end{subfigure}%
	~
	\begin{subfigure}[t]{0.32\textwidth}
		\centering
		\includegraphics[width=\textwidth, height=0.8\textwidth]{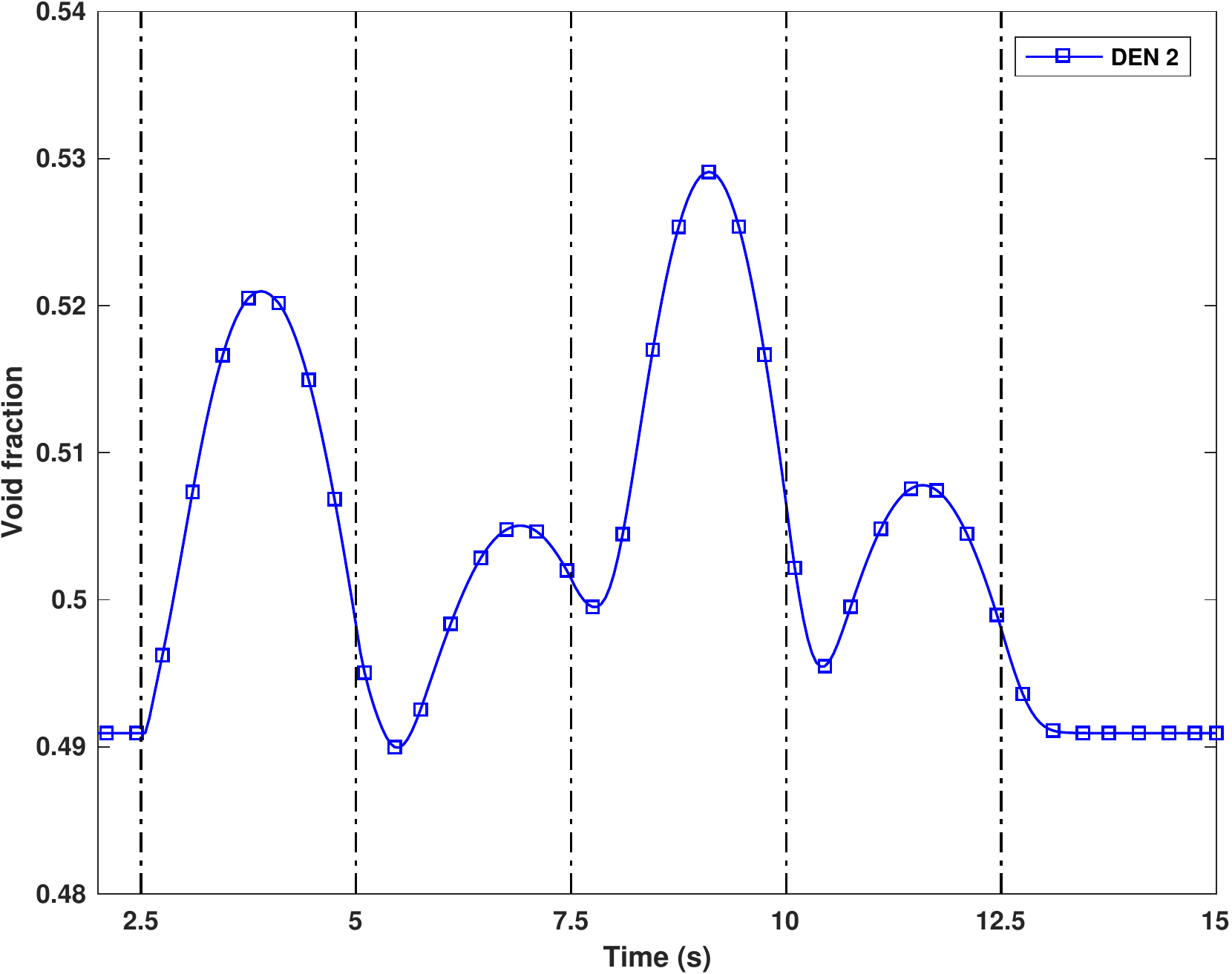}
		\caption{DEN 2}
	\end{subfigure}%
	~
	\begin{subfigure}[t]{0.32\textwidth}
		\centering
		\includegraphics[width=\textwidth, height=0.8\textwidth]{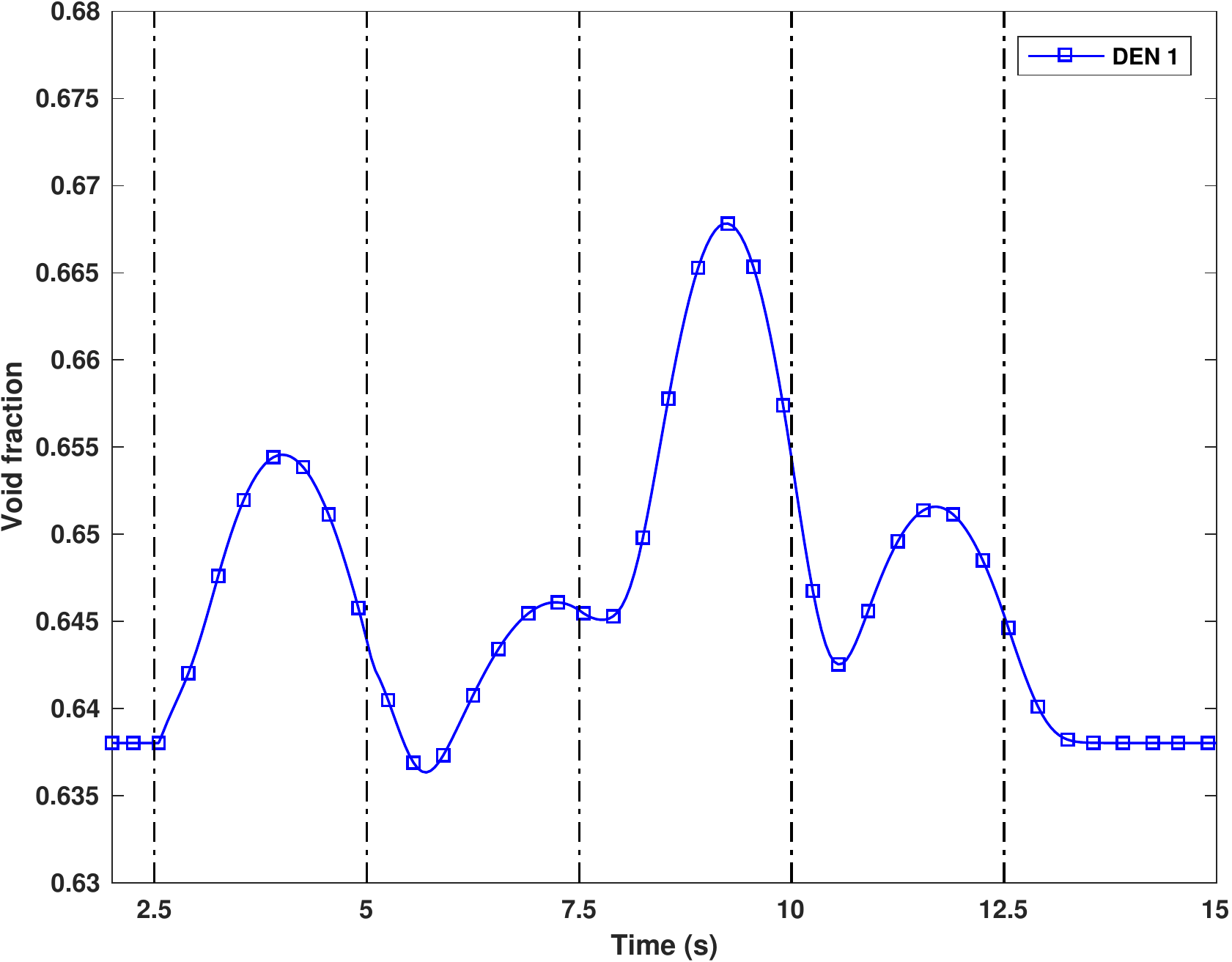}
		\caption{DEN 1}
	\end{subfigure}%
	\caption{Transient void fraction at 3 locations.} \label{bfbf-psedo-transient-forward-voidf}
\end{figure}

One transient test is designed for verification purposes. The test contains 4 transient period: outlet pressure ($p_{outlet}$) decrease transient (T1), inlet liquid temperature ($T_{l, inlet}$) increase transient (T2), inlet liquid velocity ($u_{l, inlet}$) decrease transient (T3), and assembly power ($Q$) increase transient (T4). These 4 periods simulate the typical boundary conditions change in the boiling channel that cause a increase in the void fraction. For simplicity reasons, the boundary conditions are changed through
\begin{equation}
f(t) =f_0 + f_t \cdot \sin\normp{\pi\frac{t - t_0}{t_1 - t_0}}, t_0 \le t \le t_1
\end{equation}
where $t_0$ denotes the start of the transient period, $t_1$ denotes the end of the transient period,  $f$ denotes the boundary condition (e.g. $p_{outlet}$), $f_0$ denotes the value of $f$ at time $t_0$, and $f_t$ is the change rate of $f$. These values are listed in \refTab{adjoint-bfbf-psedo-transient}. The transient void fraction at 3 locations is shown in \refFig{bfbf-psedo-transient-forward-voidf}.  Both the forward and adjoint simulations will be performed with $\Delta t = 0.05$ s and $\mathrm{N} = 48$. 

\begin{table}[!htb]
	\caption{Parameters for the transient test cases}\label{adjoint-bfbf-psedo-transient}
	\centering
	\begin{tabular}{c|ccccc}
		\hline
		Test period & $t_0$  & $t_1$   & $f$            & $f_0$      & $f_t$    \\ \hline
		T1          & 2.5 s  & 5.0 s   & $p_{outlet}$   & 7.12 MPa   & -0.2 MPa \\
		T2          & 5.0 s  & 7.5 s   & $T_{l, inlet}$ & 554.2 K    & 1.0 K    \\
		T3          & 7.5 s  & 10.0 s  & $u_{l, inlet}$ & 2.069 m/s  & -0.25 m/s \\
		T4          & 10.5 s & 12.5 s  & $Q$            & 4.53 MW    & 0.25 MW   \\
		\hline
	\end{tabular}
\end{table}

\subsection{Input parameters}
The input parameters should be chosen according to the specific purposes. For typical boiling problems, boundary conditions are usually specified by inlet liquid temperature, inlet liquid velocity, and outlet pressure; physical models are specified by net vapor generation rate ($\Gamma_{ig}, \Gamma_{w}$), interfacial friction ($f_i$), wall friction ($f_{wl}, f_{wg}$), interfacial heat flux ($Q_{il}, Q_{ig}$), and wall heat flux ($Q_{wl}, Q_{wg}$). Many of these physical models are correlated, e.g. $\Gamma_{ig}$, $Q_{il}$, and $ Q_{ig}$ are correlated through the interfacial heat transfer coefficients ($H_{il}, H_{ig}$). Excluding the wall heat flux, which is mainly affected by the assembly power, there are 5 independent physical models that worth studying in details, including $f_i $,  $f_{wl}$,  $f_{wg}$, $H_{il}$, and $H_{ig}$. In case of the subcooled boiling, the physical model for predicting the critical enthalpy ($h_{cr}$) is worth studying, as it specifies the transition from single-phase to two-phase flow.

Sensitivity of the void fraction ($\alpha_g$) to 11 parameters will be studied. These 11 parameters are 
\begin{equation}\label{bfbt-input-parameters}
\vects{\omega} = \normp{
p_{outlet},  T_{l, inlet},  u_{l, inlet}, Q, D_h , h_{cr} ,  f_i , f_{wl} , f_{wg} , H_{il} , H_{ig}}
\end{equation}
where $D_h$ is the hydraulic diameter, which represents the typical geometry parameter. A perturbation is required to calculate the coefficient matrices and vectors. The 4 boundary conditions are perturbed by the change rate, i.e.
\begin{equation}
f_t = f_t + \varepsilon
\end{equation}
where $\varepsilon$ denotes the perturbation. The hydraulic diameter is perturbed by
\begin{equation}
D_h    =  D_h + \varepsilon
\end{equation}
The critical enthalpy is perturbed by
\begin{equation}
\mathrm{St}_{cr} = \normp{1 + \varepsilon}\cdot\mathrm{St}_{cr, 0}, \mathrm{Nu}_{cr} = \normp{1 + \varepsilon}\cdot\mathrm{Nu}_{cr, 0}
\end{equation}
where $\mathrm{St}_{cr, 0}$ and $\mathrm{Nu}_{cr, 0}$ are the nominal values used by \refEq{Saha-Zuber-model}. The remaining 5 physical model parameters are perturbed by,  
\begin{equation}
\omega    = \normp{1 + \varepsilon } \omega_{0}
\end{equation}
Perturbation in this form is mainly a numerical compromise to avoid messing with the existing closure correlations; however, it does not affect the application of the adjoint method to calculate the sensitivity. Note that $\omega_{0}$  is obtained directly from the existing closure correlations and remains unchanged.

\subsection{Results}
The response of interest is the void fraction at 3 locations (DEN 3, DEN 2, and DEN 1) and 50 time steps starting from 2.5 to 15.0 s. Sensitivities to all input parameters are calculated with perturbation sensitivity analysis (PS), continuous adjoint sensitivity analysis (CAS), and discrete adjoint sensitivity analysis (DAS). A large volume of the sensitivity result is generated. For brevity reasons, only a few series of the result will be shown quantitatively. The sensitive input parameters are the change rate of 4 boundary conditions.  

\refFig{bfbf-psedo-psedo-adjoint-variable-phi-a}, \refFig{bfbf-psedo-psedo-adjoint-variable-phi-b}, and \refFig{bfbf-psedo-psedo-adjoint-variable-phi-c} show examples of the time and space  dependent adjoint variables given by both CAS and DAS schemes. The void fraction at 3 locations and at $t_d = 8.75$ s is the response of interest. The profile of the adjoint variables from CAS and DAS schemes are similar, which means that the CAS and DAS are consistent with each other. Vertically, it is seen that the adjoint variable is non-zero for $x > x_d$ but negligible for $x > x_d$, which reflects that the response is mainly affected by forward solution in its upwind side. Horizontally, it is seen that when the time $t$ decreases, the adjoint variable decreases rapidly to 0, which reflects that the response is mainly affected by its prior but close state.  

\begin{figure}[!htbp]
	\centering
	\begin{subfigure}[t]{0.5\textwidth}
		\includegraphics[width=\textwidth]{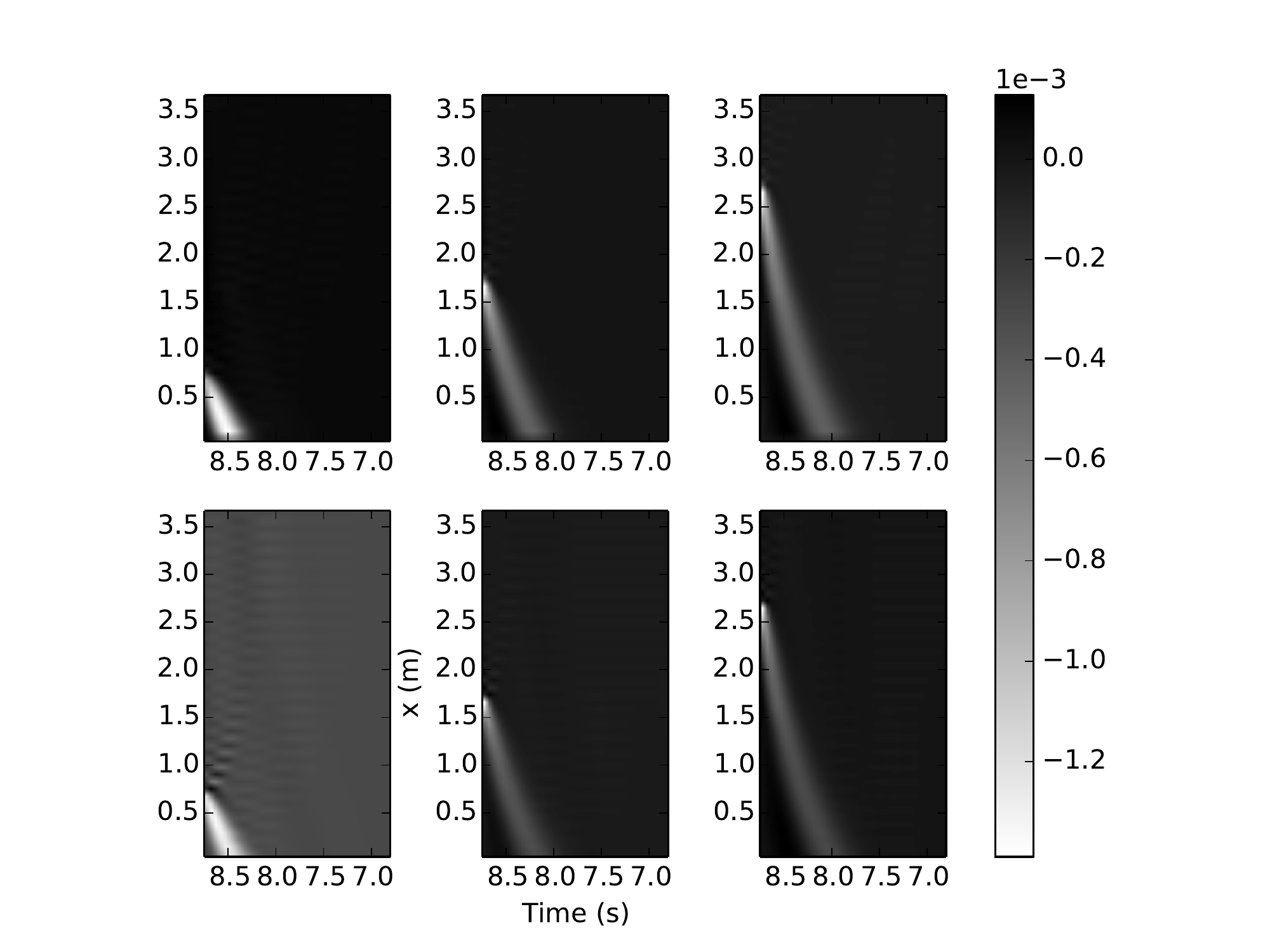}
		\caption{$\phi_1$ - Liquid mass equation}
	\end{subfigure}%
	\hspace{-1cm}
	\begin{subfigure}[t]{0.5\textwidth}
		\includegraphics[width=\textwidth]{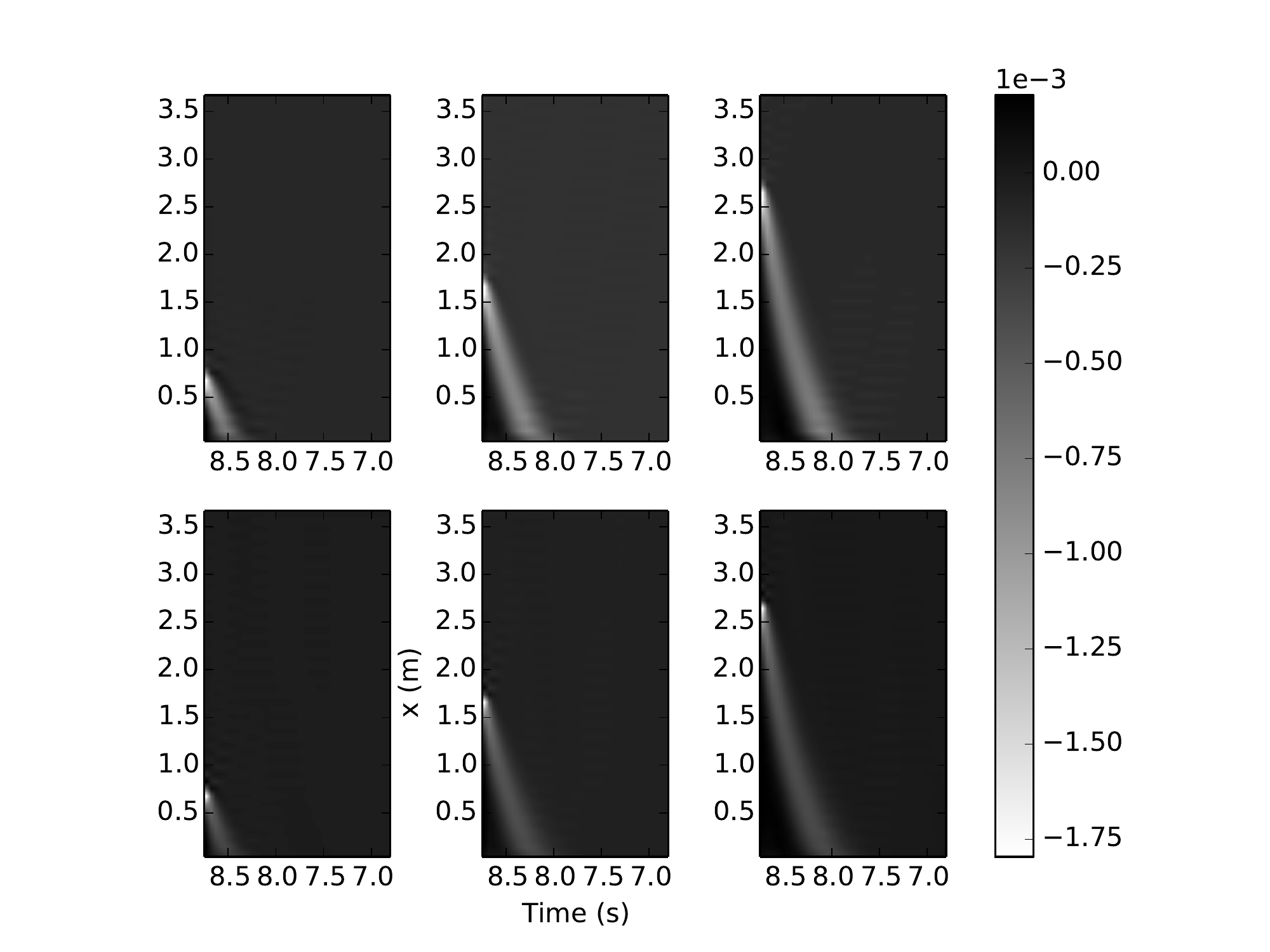}
		\caption{$\phi_4$ - Gas mass equation}
	\end{subfigure}%
	\caption{Adjoint solution given by DAS (upper row) and CAS (lower row) for void fraction at DEN 3 (left column), DEN 2 (middle column), and DEN 1 (right column). The time of interest is $t_d = 8.75$ s.} \label{bfbf-psedo-psedo-adjoint-variable-phi-a}
\end{figure}

\begin{figure}[!htbp]
	\centering
	\begin{subfigure}[t]{0.5\textwidth}
		\includegraphics[width=\textwidth]{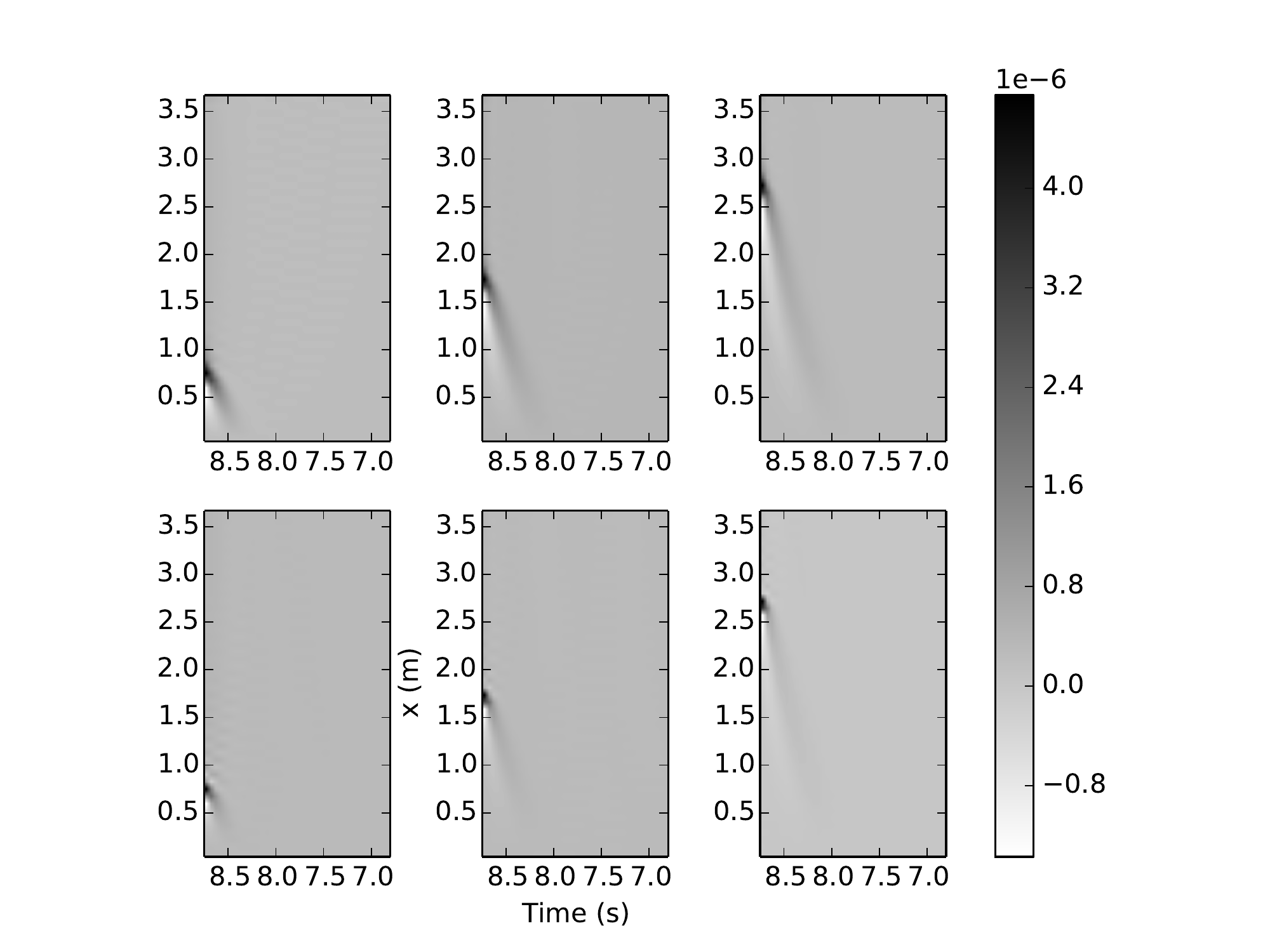}
		\caption{$\phi_2$ - Liquid momentum equation}
	\end{subfigure}%
	\hspace{-1cm}
	\begin{subfigure}[t]{0.5\textwidth}
		\includegraphics[width=\textwidth]{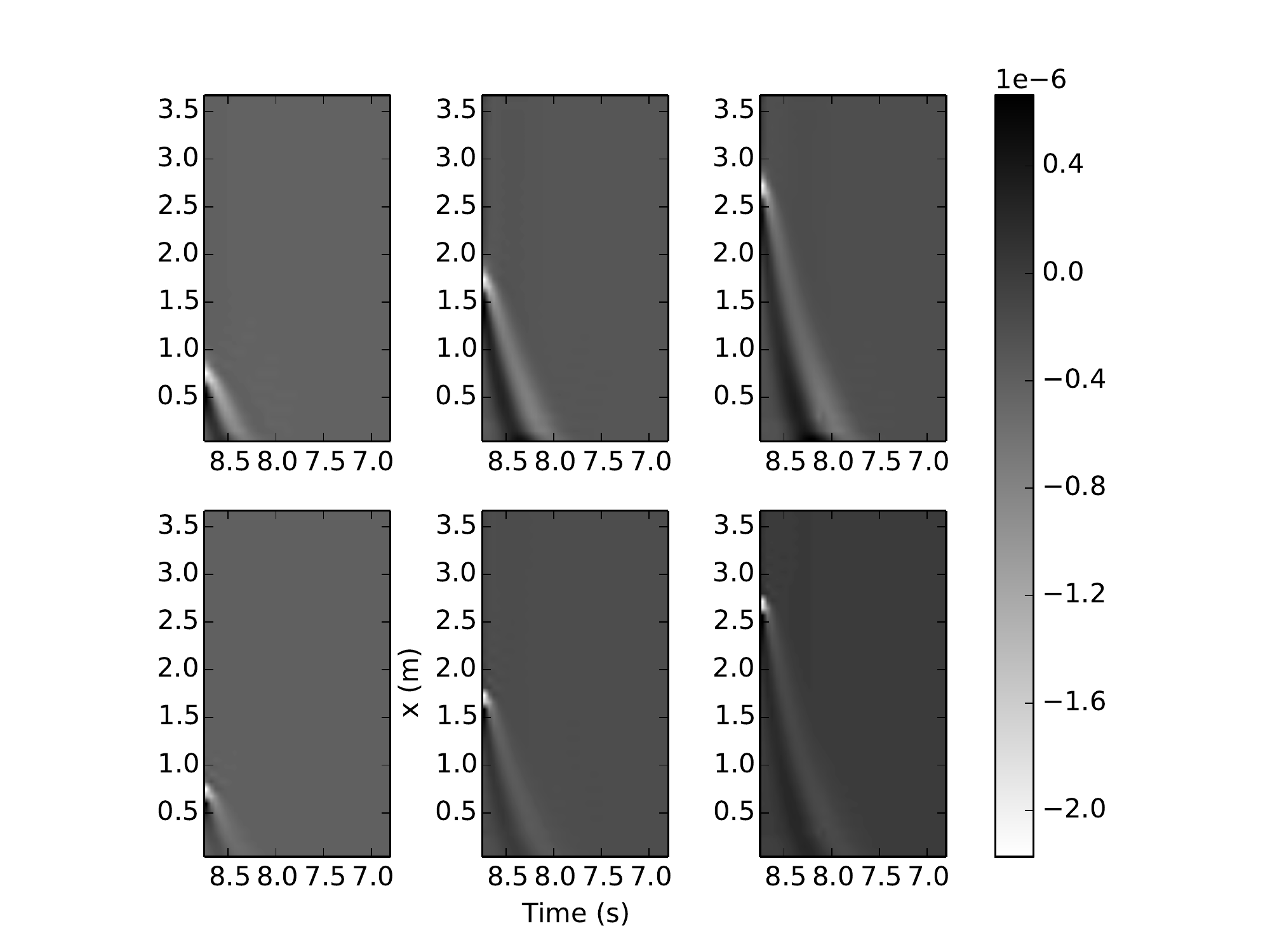}
		\caption{$\phi_5$ - Gas momentum equation}
	\end{subfigure}%
	\caption{Continued: adjoint solution given by DAS (upper row) and CAS (lower row) for void fraction at DEN 3 (left column), DEN 2 (middle column), and DEN 1 (right column). The time of interest is $t_d = 8.75$ s.} \label{bfbf-psedo-psedo-adjoint-variable-phi-b}
\end{figure}

\begin{figure}[!htbp]
	\centering
	\begin{subfigure}[t]{0.5\textwidth}
		\includegraphics[width=\textwidth]{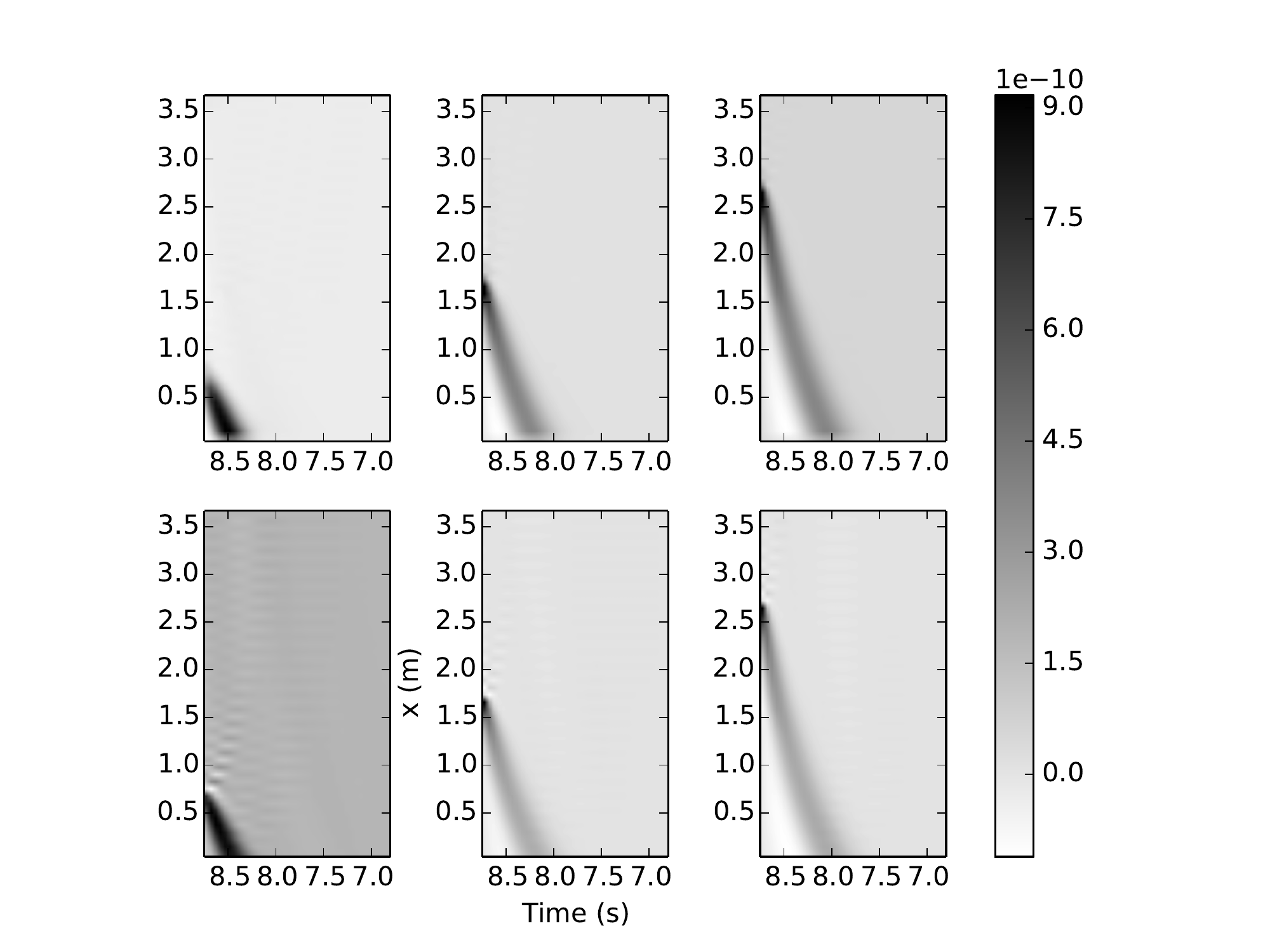}
		\caption{$\phi_3$ - Liquid energy equation}
	\end{subfigure}%
	\hspace{-1cm}
	\begin{subfigure}[t]{0.5\textwidth}
		\includegraphics[width=\textwidth]{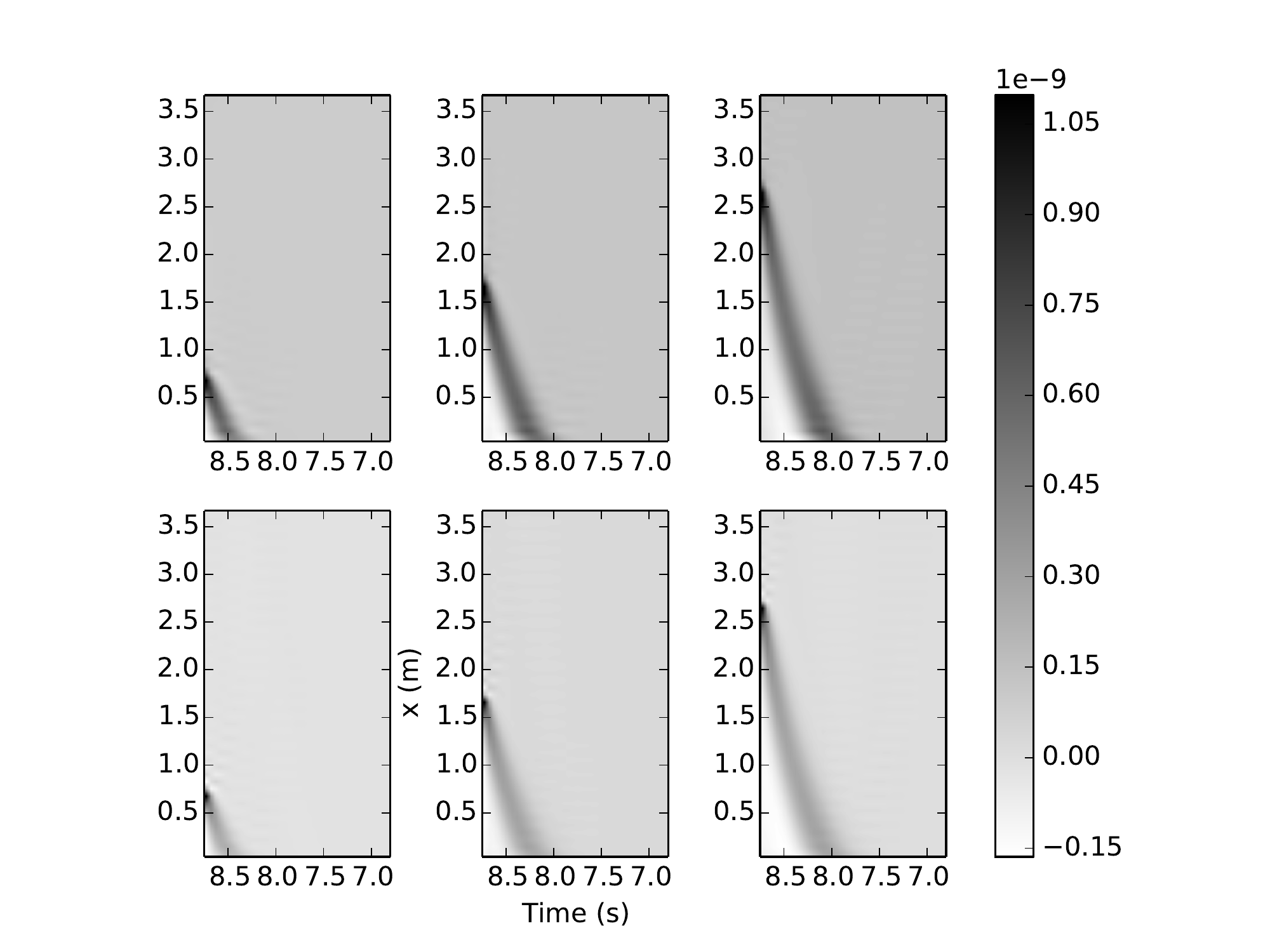}
		\caption{$\phi_6$ - Gas energy equation}
	\end{subfigure}%
	\caption{Continued: adjoint solution given by DAS (upper row) and CAS (lower row) for void fraction at DEN 3 (left column), DEN 2 (middle column), and DEN 1 (right column). The time of interest is $t_d = 8.75$ s.} \label{bfbf-psedo-psedo-adjoint-variable-phi-c}
\end{figure}

For clarity reasons, let's denote adjoint variable with
\begin{equation}\label{adjoint-variable-notation}
\vects{\phi} = \vects{\phi}\normb{\normp{x, t}\rightarrow \normp{x_d, t_d}}
\end{equation}
In each time step, there are 6 components in the adjoint vector in each location. The component of the adjoint vector represents the effect of the corresponding governing equation, e.g. $\phi_1$ for the liquid mass equation, to the response. The notation $\normp{x, t}\rightarrow \normp{x_d, t_d}$ in \refEq{adjoint-variable-notation} represents the effect of the system at $\normp{x, t}$ to the response of interest at $\normp{x_d, t_d}$.

\begin{figure}[!htbp]
	\centering
	\includegraphics[width=0.5\textwidth]{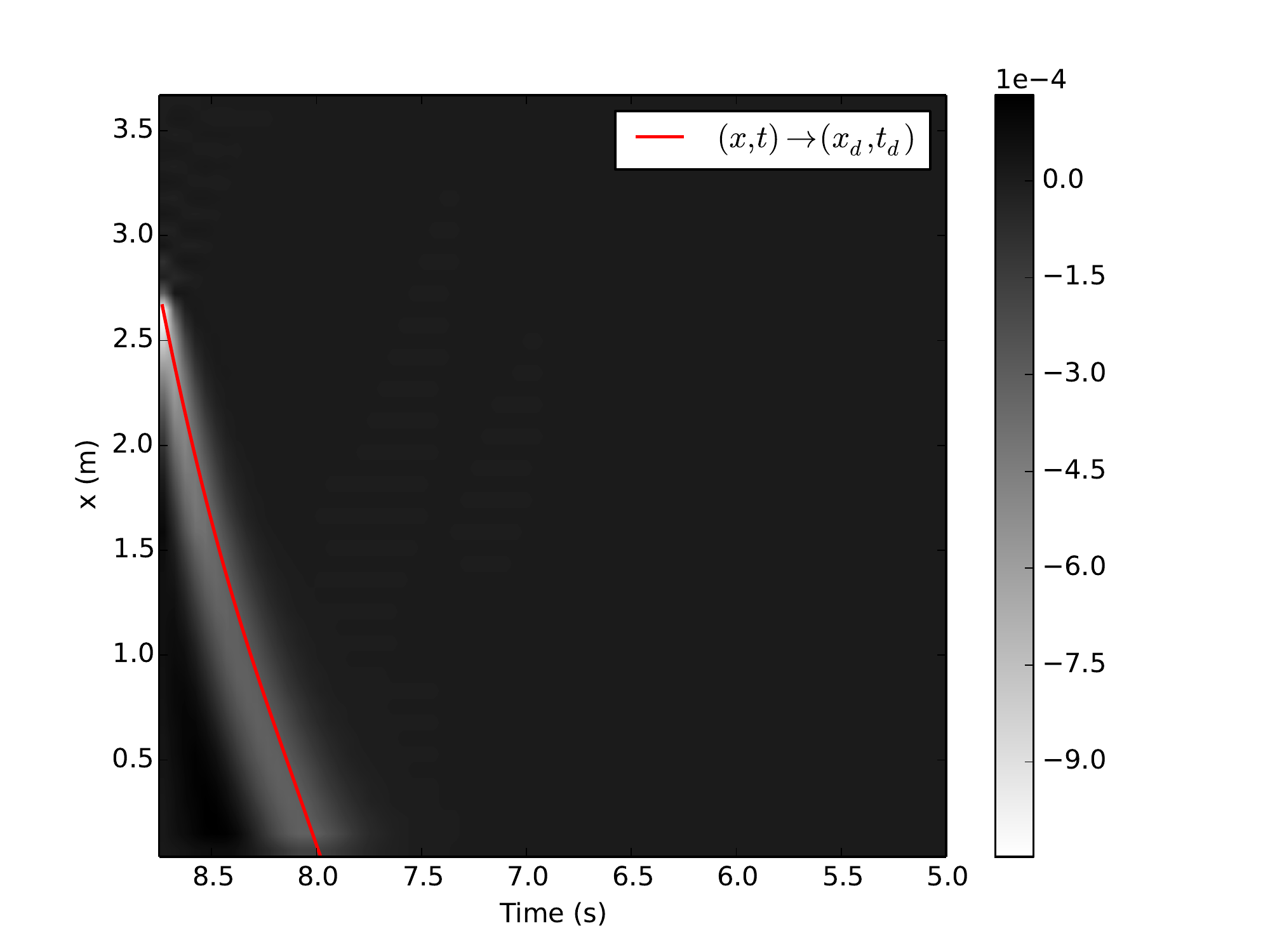}
	\caption{Adjoint solution ($\phi_1$) given by DAS for void fraction at DEN 3 ($x_d = 2.730$ m) and $t_d = 8.75$ s. } \label{bfbt-psedo-adjoint-variable-p1-with-bc}
\end{figure}

Let $t_{\mathrm{eff}}\normb{\normp{x, t}\rightarrow \normp{x_d, t_d}}$ be the time period when $\vects{\phi}\normb{\normp{x, t}\rightarrow \normp{x_d, t_d}}$ is the biggest. There are 6 characteristic waves in the two-phase system, including both acoustic waves and the convection waves. Because the time step is too large to capture the the acoustic waves, the remaining important wave is the convection of liquid enthalpy, which has a wave speed of the liquid velocity. Thus, $t_{\mathrm{eff}}$ can be approximately by
\begin{equation}\label{adjoint-time-prediction}
t_{\mathrm{eff}}\normb{\normp{x, t}\rightarrow \normp{x_d, t_d}} \approx \frac{2\normp{x_d - x}}{u_{l}(x, t) + u_{l}(x_d, t_d)}
\end{equation} 
\refEq{adjoint-time-prediction} gives a curve in $x-t$ space that represents the location of the maximum adjoint solution, see \refFig{bfbt-psedo-adjoint-variable-p1-with-bc}. The good agreement of this curve verifies the current adjoint sensitivity analysis framework. 

\begin{figure}[!htbp]
	\centering
	\begin{subfigure}[t]{0.32\textwidth}
		\includegraphics[width=\textwidth]{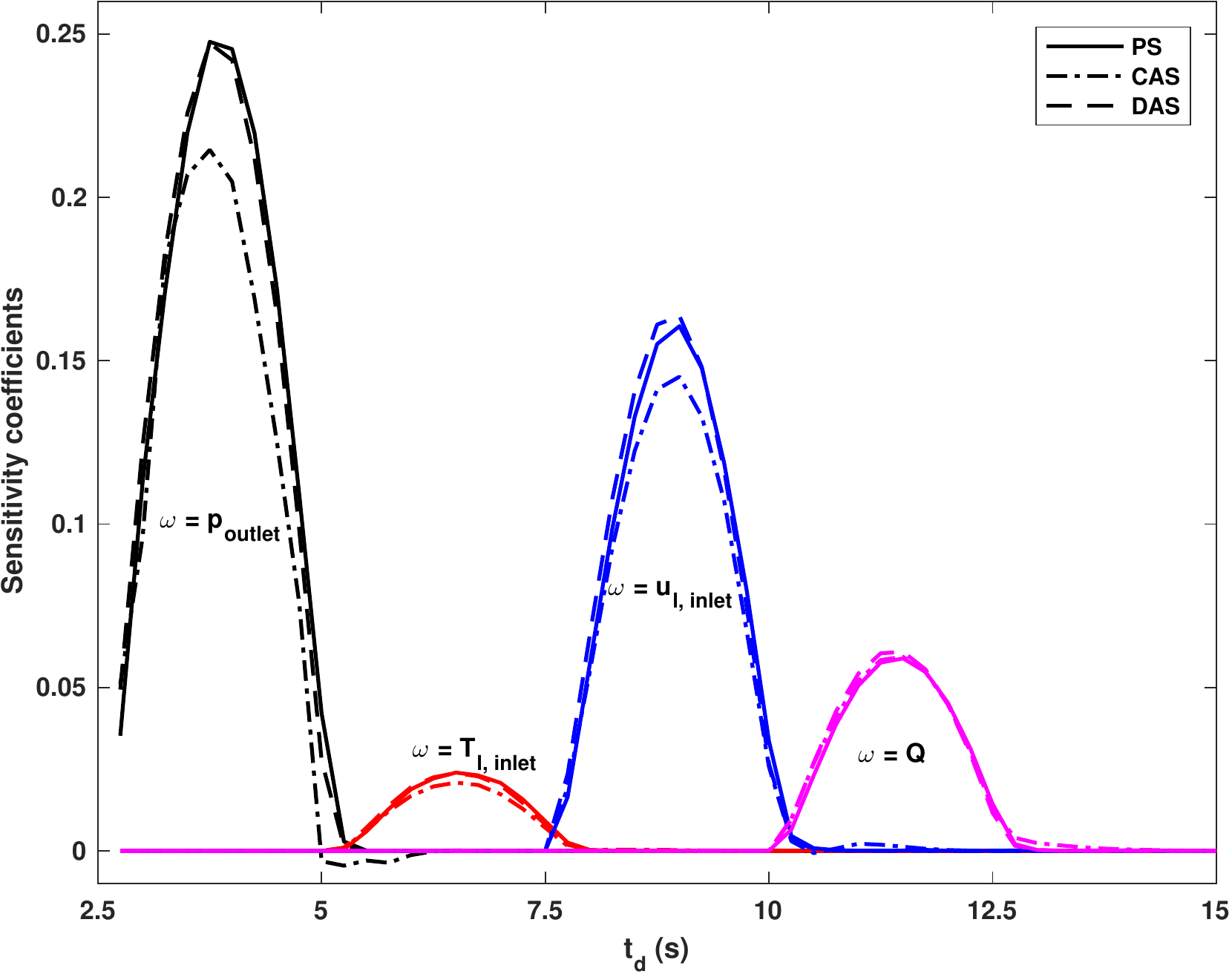}
		\caption{DEN 3}
	\end{subfigure}%
	~
	\begin{subfigure}[t]{0.32\textwidth}
		\includegraphics[width=\textwidth]{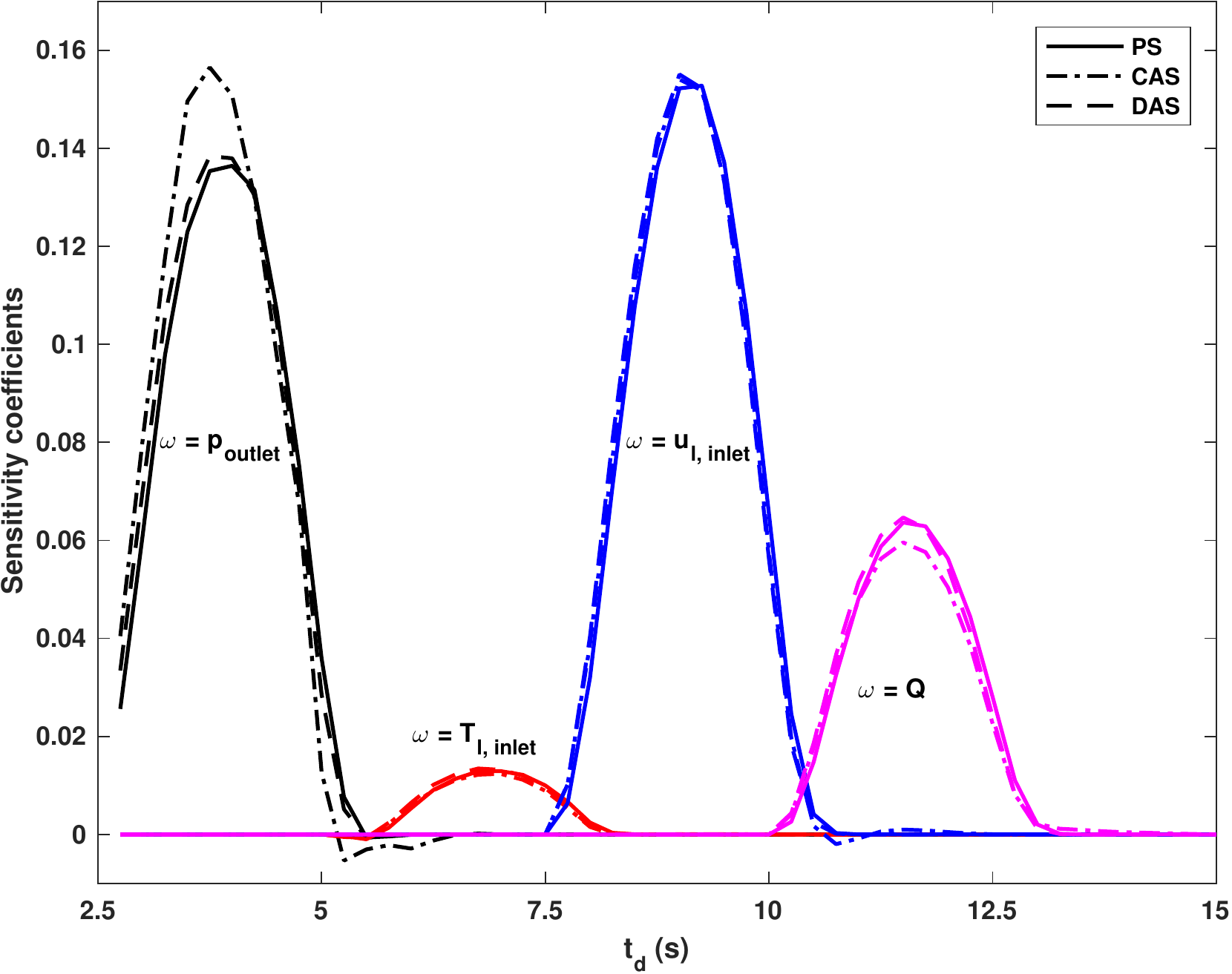}
		\caption{DEN 2}
	\end{subfigure}%
	~
	\begin{subfigure}[t]{0.32\textwidth}
		\includegraphics[width=\textwidth]{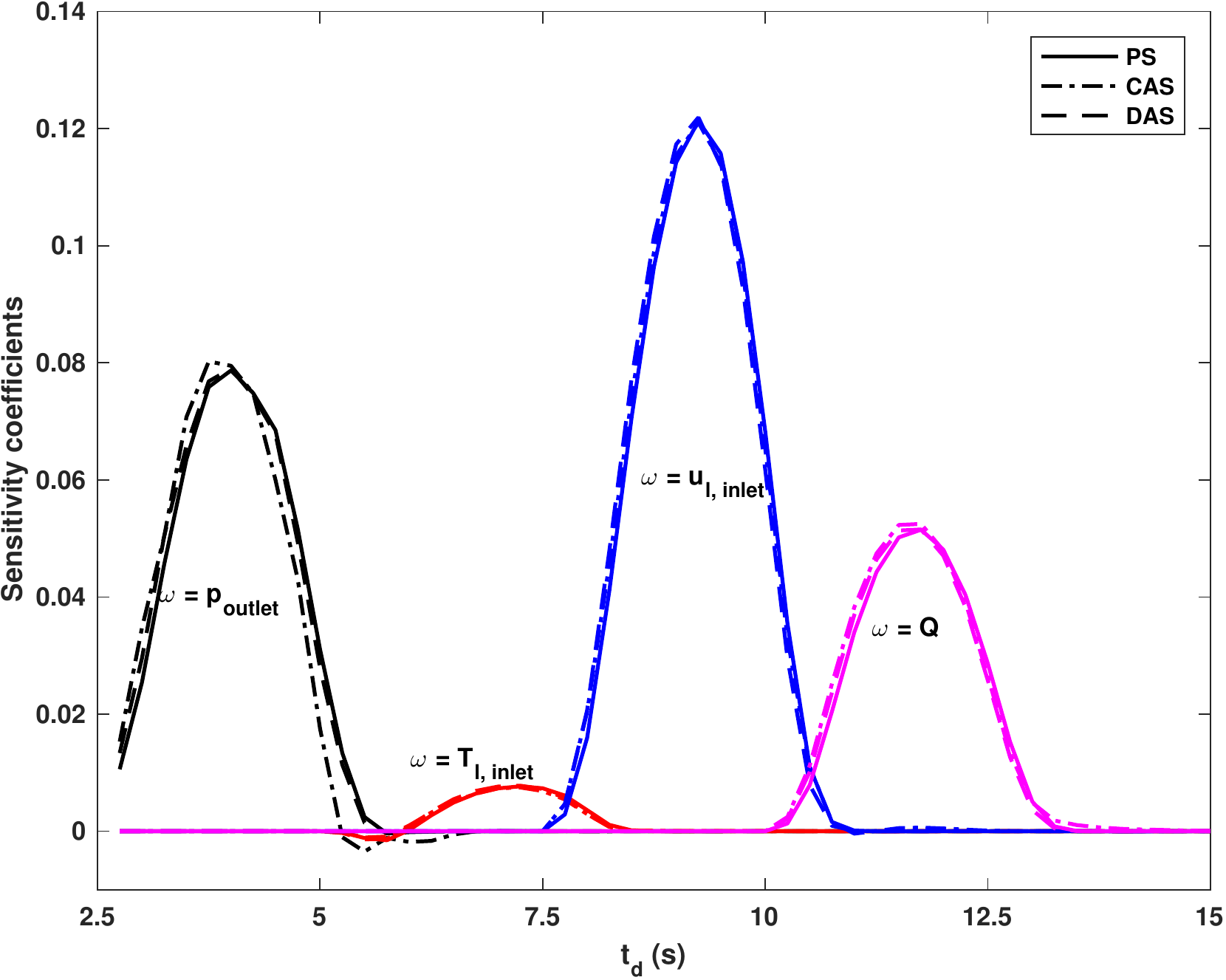}
		\caption{DEN 1}
	\end{subfigure}%
	\caption{Sensitivity coefficients of void fraction at 3 locations to the change rate of 4 boundary conditions.} \label{bfbf-psedo-psedo-adjoint-sa-voidf-S}
\end{figure}

\begin{table}[!htbp]
	\centering
	\caption{Comparison of sensitivities from CAS and DAS with sensitivities from PS}\label{SS-BFBT-SA-Error-T1-a}
	\resizebox{\textwidth}{!}{
		\begin{tabular}{c|rrr|rrr|rrr}
			\hline
			&  \multicolumn{3}{|c|}{DEN 3} &\multicolumn{3}{|c|}{DEN 2}&\multicolumn{3}{|c}{DEN 1} \\ \hline
			$t_d$:(s) & PS & CAS & DAS  & PS & CAS & DAS & PS & CAS & DAS \\ \hline
			&\multicolumn{9}{c}{$\omega = p_{outlet}$} \\ \hline
			3.00	&	1.13E-01	&	9.60E-02	&	1.24E-01	&	6.10E-02	&	8.05E-02	&	7.02E-02	&	2.54E-02	&	3.47E-02	&	2.95E-02	\\
			3.50	&	2.20E-01	&	2.07E-01	&	2.26E-01	&	1.23E-01	&	1.50E-01	&	1.28E-01	&	6.35E-02	&	7.08E-02	&	6.58E-02	\\
			4.00	&	2.45E-01	&	2.05E-01	&	2.42E-01	&	1.36E-01	&	1.51E-01	&	1.38E-01	&	7.87E-02	&	7.95E-02	&	7.90E-02	\\
			4.50	&	1.72E-01	&	1.25E-01	&	1.64E-01	&	1.07E-01	&	9.75E-02	&	1.04E-01	&	6.85E-02	&	6.00E-02	&	6.78E-02	\\ \hline
			&\multicolumn{9}{c}{$\omega = T_{l, inlet}$} \\ \hline
			5.50	&	5.90E-03	&	6.22E-03	&	7.02E-03	&	-9.76E-04	&	-5.86E-04	&	-6.88E-04	&	-1.16E-03	&	-1.07E-03	&	-1.38E-03	\\
			6.00	&	1.89E-02	&	1.68E-02	&	1.95E-02	&	5.32E-03	&	5.88E-03	&	6.45E-03	&	1.92E-04	&	9.78E-04	&	7.65E-04	\\
			6.50	&	2.39E-02	&	2.09E-02	&	2.38E-02	&	1.14E-02	&	1.11E-02	&	1.23E-02	&	4.90E-03	&	5.06E-03	&	5.47E-03	\\
			7.00	&	2.08E-02	&	1.73E-02	&	2.01E-02	&	1.29E-02	&	1.24E-02	&	1.32E-02	&	7.33E-03	&	7.33E-03	&	7.64E-03	\\ \hline
			&\multicolumn{9}{c}{$\omega = u_{l, inlet}$} \\ \hline
			8.00	&	5.75E-02	&	5.52E-02	&	6.65E-02	&	3.20E-02	&	3.89E-02	&	4.04E-02	&	1.59E-02	&	2.04E-02	&	2.07E-02	\\
			8.50	&	1.33E-01	&	1.22E-01	&	1.41E-01	&	1.08E-01	&	1.13E-01	&	1.16E-01	&	7.08E-02	&	7.41E-02	&	7.74E-02	\\
			9.00	&	1.61E-01	&	1.45E-01	&	1.64E-01	&	1.52E-01	&	1.54E-01	&	1.55E-01	&	1.14E-01	&	1.15E-01	&	1.17E-01	\\
			9.50	&	1.19E-01	&	1.07E-01	&	1.16E-01	&	1.37E-01	&	1.33E-01	&	1.33E-01	&	1.16E-01	&	1.14E-01	&	1.14E-01	\\ \hline
			&\multicolumn{9}{c}{$\omega = Q$} \\ \hline
			10.50	&	2.30E-02	&	2.67E-02	&	2.70E-02	&	1.48E-02	&	1.78E-02	&	1.87E-02	&	7.88E-03	&	1.12E-02	&	1.01E-02	\\
			11.00	&	5.06E-02	&	5.21E-02	&	5.42E-02	&	4.80E-02	&	4.78E-02	&	5.15E-02	&	3.40E-02	&	3.82E-02	&	3.69E-02	\\
			11.50	&	5.89E-02	&	5.94E-02	&	6.09E-02	&	6.37E-02	&	5.96E-02	&	6.47E-02	&	5.02E-02	&	5.23E-02	&	5.14E-02	\\
			12.00	&	4.53E-02	&	4.48E-02	&	4.48E-02	&	5.63E-02	&	5.04E-02	&	5.45E-02	&	4.81E-02	&	4.80E-02	&	4.71E-02	\\ \hline
	\end{tabular}}
\end{table}

\begin{table}[!htbp]
	\centering
	\caption{Comparison of sensitivities from CAS and DAS with sensitivities from PS: continued }\label{SS-BFBT-SA-Error-T1-b}
	\resizebox{\textwidth}{!}{
		\begin{tabular}{c|rrr|rrr|rrr}
			\hline
			&  \multicolumn{3}{|c|}{DEN 3} &\multicolumn{3}{|c|}{DEN 2}&\multicolumn{3}{|c}{DEN 1} \\ \hline
			$t_d$:(s) & PS & CAS & DAS  & PS & CAS & DAS & PS & CAS & DAS \\ \hline
			&\multicolumn{9}{c}{$\omega = D_{h}$} \\ \hline
			3.75	&	-1.96E-01	&	-2.50E-01	&	-2.06E-01	&	-2.57E-01	&	-2.43E-01	&	-2.58E-01	&	-2.19E-01	&	-2.26E-01	&	-2.20E-01	\\
			6.25	&	-1.93E-01	&	-2.59E-01	&	-2.05E-01	&	-2.80E-01	&	-2.54E-01	&	-2.82E-01	&	-2.36E-01	&	-2.44E-01	&	-2.38E-01	\\
			8.75	&	-2.08E-01	&	-2.75E-01	&	-2.23E-01	&	-2.78E-01	&	-2.64E-01	&	-2.81E-01	&	-2.31E-01	&	-2.46E-01	&	-2.33E-01	\\
			11.25	&	-1.97E-01	&	-2.38E-01	&	-2.13E-01	&	-2.83E-01	&	-2.69E-01	&	-2.86E-01	&	-2.35E-01	&	-2.48E-01	&	-2.37E-01	\\
			13.75	&	-1.88E-01	&	-2.16E-01	&	-2.05E-01	&	-2.88E-01	&	-2.72E-01	&	-2.89E-01	&	-2.38E-01	&	-2.52E-01	&	-2.40E-01	\\ \hline
			&\multicolumn{9}{c}{$\omega = h_{cr}$} \\ \hline
			3.75	&	9.23E-03	&	8.96E-03	&	9.60E-03	&	1.16E-02	&	1.20E-02	&	1.16E-02	&	1.01E-02	&	1.04E-02	&	1.01E-02	\\
			6.25	&	8.73E-03	&	8.60E-03	&	9.12E-03	&	1.15E-02	&	1.18E-02	&	1.16E-02	&	1.02E-02	&	1.05E-02	&	1.03E-02	\\
			8.75	&	1.01E-02	&	9.96E-03	&	1.06E-02	&	1.28E-02	&	1.30E-02	&	1.29E-02	&	1.08E-02	&	1.10E-02	&	1.09E-02	\\
			11.25	&	8.34E-03	&	8.26E-03	&	8.71E-03	&	1.18E-02	&	1.20E-02	&	1.19E-02	&	1.03E-02	&	1.05E-02	&	1.04E-02	\\
			13.75	&	7.91E-03	&	7.82E-03	&	8.36E-03	&	1.20E-02	&	1.22E-02	&	1.20E-02	&	1.06E-02	&	1.08E-02	&	1.07E-02	\\ \hline
			&\multicolumn{9}{c}{$\omega = f_{i}$} \\ \hline
			3.75	&	-1.08E-02	&	-8.37E-03	&	-1.09E-02	&	-8.53E-03	&	-8.44E-03	&	-8.54E-03	&	-6.81E-03	&	-6.29E-03	&	-6.77E-03	\\
			6.25	&	-9.66E-03	&	-8.96E-03	&	-9.98E-03	&	-8.49E-03	&	-7.41E-03	&	-8.55E-03	&	-6.85E-03	&	-5.64E-03	&	-6.83E-03	\\
			8.75	&	-9.17E-03	&	-7.18E-03	&	-9.13E-03	&	-8.08E-03	&	-8.22E-03	&	-8.11E-03	&	-6.88E-03	&	-6.33E-03	&	-6.88E-03	\\
			11.25	&	-1.01E-02	&	-7.25E-03	&	-1.00E-02	&	-8.39E-03	&	-8.28E-03	&	-8.48E-03	&	-6.73E-03	&	-6.05E-03	&	-6.76E-03	\\
			13.75	&	-9.74E-03	&	-7.19E-03	&	-9.68E-03	&	-8.36E-03	&	-8.28E-03	&	-8.46E-03	&	-6.75E-03	&	-6.05E-03	&	-6.79E-03	\\ 
			\hline
	\end{tabular}}
\end{table}

\begin{table}[!htbp]
	\centering
	\caption{Comparison of sensitivities from CAS and DAS with sensitivities from PS: continued }\label{SS-BFBT-SA-Error-T1-c}
	\resizebox{\textwidth}{!}{
		\begin{tabular}{c|rrr|rrr|rrr}
			\hline
			&  \multicolumn{3}{|c|}{DEN 3} &\multicolumn{3}{|c|}{DEN 2}&\multicolumn{3}{|c}{DEN 1} \\ \hline
			$t_d$:(s) & PS & CAS & DAS  & PS & CAS & DAS & PS & CAS & DAS \\ \hline
			&\multicolumn{9}{c}{$\omega = f_{wl}$} \\ \hline
			3.75	&	-1.46E-03	&	-1.10E-03	&	-1.43E-03	&	-4.55E-04	&	-4.89E-04	&	-4.73E-04	&	1.20E-04	&	2.47E-04	&	1.16E-04	\\
			6.25	&	-1.18E-03	&	-1.16E-03	&	-1.24E-03	&	-4.49E-04	&	-3.08E-04	&	-4.60E-04	&	1.05E-04	&	3.56E-04	&	1.18E-04	\\
			8.75	&	-1.28E-03	&	-9.67E-04	&	-1.23E-03	&	-2.99E-04	&	-3.96E-04	&	-3.21E-04	&	2.38E-04	&	3.66E-04	&	2.30E-04	\\
			11.25	&	-1.38E-03	&	-9.42E-04	&	-1.36E-03	&	-4.31E-04	&	-4.90E-04	&	-4.72E-04	&	1.60E-04	&	2.94E-04	&	1.35E-04	\\
			13.75	&	-1.27E-03	&	-8.90E-04	&	-1.24E-03	&	-4.04E-04	&	-4.56E-04	&	-4.37E-04	&	1.61E-04	&	3.08E-04	&	1.50E-04	\\ \hline
			&\multicolumn{9}{c}{$\omega = f_{wg}$} \\ \hline
			3.75	&	1.74E-03	&	7.63E-04	&	1.86E-03	&	6.60E-04	&	1.30E-03	&	8.69E-04	&	-2.41E-03	&	-2.29E-03	&	-2.34E-03	\\
			6.25	&	1.77E-03	&	1.65E-03	&	1.89E-03	&	2.49E-03	&	3.09E-03	&	2.67E-03	&	-2.66E-04	&	-8.59E-05	&	-1.70E-04	\\
			8.75	&	2.07E-03	&	2.08E-03	&	2.38E-03	&	2.83E-03	&	3.16E-03	&	3.00E-03	&	1.26E-04	&	1.54E-04	&	1.96E-04	\\
			11.25	&	1.81E-03	&	1.93E-03	&	2.10E-03	&	2.77E-03	&	3.18E-03	&	2.89E-03	&	-1.10E-04	&	-7.91E-05	&	-6.44E-05	\\
			13.75	&	1.64E-03	&	1.82E-03	&	1.96E-03	&	2.85E-03	&	3.29E-03	&	2.97E-03	&	-1.28E-04	&	-8.33E-05	&	-5.86E-05	\\ \hline
			&\multicolumn{9}{c}{$\omega = H_{il}$} \\ \hline
			3.75	&	3.45E-05	&	1.37E-04	&	4.16E-05	&	1.73E-04	&	1.31E-04	&	1.71E-04	&	2.80E-04	&	2.77E-04	&	2.76E-04	\\
			6.25	&	1.09E-04	&	5.99E-05	&	1.47E-05	&	1.16E-04	&	4.57E-05	&	9.39E-05	&	1.58E-04	&	1.64E-04	&	1.71E-04	\\
			8.75	&	1.27E-05	&	7.00E-05	&	1.58E-05	&	8.92E-05	&	6.56E-05	&	8.66E-05	&	1.56E-04	&	1.64E-04	&	1.54E-04	\\
			11.25	&	1.27E-04	&	3.62E-05	&	1.18E-05	&	9.15E-05	&	6.57E-05	&	9.15E-05	&	1.71E-04	&	1.75E-04	&	1.68E-04	\\
			13.75	&	7.37E-06	&	2.13E-05	&	1.03E-05	&	8.72E-05	&	6.31E-05	&	8.82E-05	&	1.61E-04	&	1.73E-04	&	1.65E-04	\\ 
			\hline
			&\multicolumn{9}{c}{$\omega = H_{ig}$} \\ \hline
			3.75	&	-3.84E-02	&	-2.40E-02	&	-3.58E-02	&	-2.87E-03	&	-8.73E-04	&	-2.67E-03	&	-1.72E-03	&	-8.46E-04	&	-1.48E-03	\\
			6.25	&	-4.70E-02	&	-3.10E-02	&	-4.45E-02	&	-4.09E-03	&	-1.43E-03	&	-4.32E-03	&	-2.33E-03	&	-1.19E-03	&	-2.32E-03	\\
			8.75	&	-5.18E-02	&	-3.14E-02	&	-4.71E-02	&	-3.25E-03	&	-8.57E-04	&	-3.27E-03	&	-1.79E-03	&	-8.31E-04	&	-1.65E-03	\\
			11.25	&	-5.86E-02	&	-3.71E-02	&	-5.34E-02	&	-3.02E-03	&	-5.66E-04	&	-3.41E-03	&	-1.00E-03	&	-4.04E-04	&	-9.71E-04	\\
			13.75	&	-6.07E-02	&	-3.90E-02	&	-5.55E-02	&	-2.55E-03	&	-3.11E-04	&	-3.04E-03	&	-3.69E-04	&	-1.17E-04	&	-3.99E-04	\\ 
			\hline
	\end{tabular}}
\end{table}

\refFig{bfbf-psedo-psedo-adjoint-sa-voidf-S} shows the comparison of sensitivities from different methods. The input parameters are the change rate of 4 boundary conditions. \refTab{SS-BFBT-SA-Error-T1-a}, \refTab{SS-BFBT-SA-Error-T1-b}, and \refTab{SS-BFBT-SA-Error-T1-c} show quantitatively the comparison of sensitivities at selected time steps from CAS and DAS to sensitivities from PS. Overall, it is found that the sensitivities from DAS and CAS match that of PS well, which verify the adjoint sensitivity analysis framework. However, it is seen that the sensitivities from CAS are not as accurate as sensitivities from DAS. This is because of two reasons. The first reason is in the solution scheme for solving the continuous adjoint equation, the central finite difference scheme. For the response function considered in this test, the source vector $\vect{Q}$ in the continuous adjoint equation behaves like a Dirac delta function, this causes a difficulty in solving the adjoint variable. The second reason in in the difficulty in obtaining correct derivatives required by the coefficient matrices/vectors. Though the DAS is computationally more expensive due to the effort for evaluating the coefficient matrices/vectors than the CAS, the DAS is numerically more robust than the CAS.

\section{Conclusion}\label{sec-five}
In this article, an adjoint sensitivity analysis framework using the continuous and discrete adjoint method is developed and verified for transient two-phase flow simulations. Both methods are applied to one transient test case in a rod bundle. It is found that both methods give physically reasonable sensitivities to different input parameter. The discrete method is closely related to and consistent with the forward (upwind) discretization scheme and is found to be more robust than the continuous method. However it is found that the continuous method has difficulties in handling problems where there are discontinuities in the source terms. This is because of the central method in discretizing the continuous adjoint equation. Compared with the continuous method, the discrete method takes more computational time because it needs to evaluate the coefficient matrices with a numerical differentiation scheme.

\section*{Acknowledgment}
The authors would like to thank the anonymous reviewers for their detailed and valuable comments to improve the quality of the paper.

\appendix
\section{Approximate eigenvalues/eigenvectors}
The Jacobian matrix $\mat{A}_c$ is
\begin{equation}\label{Eq-App-17}
\mat{A}_c = \scalebox{.9}{$\begin{pmatrix}
	0 & 1 & 0 & 0 & 0 & 0 \\
	-u_l^2+\beta_l c_{l}^{h} & 2u_l-\beta_l c_{l}^{u} & \beta_l c_{l}^{1} & \sigma_l c_{g}^{h} & -\sigma_l c_{g}^{u} & \sigma_l c_{g}^{1} \\
	-u_l H_l +  u_l\beta_l c_{l}^{h} & H_l -u_l\beta_l c_{l}^{u} & u_l+u_l\beta_l c_{l}^{1} & \sigma_l u_l c_{g}^{h} & -\sigma_l u_l c_{g}^{u} & \sigma_l u_l c_{g}^{1} \\
	0 & 0 & 0 & 0 & 1 & 0 \\
	\sigma_g c_{l}^{h} & -\sigma_g c_{l}^{u} & \sigma_g c_{l}^{1} & -u_g^2+\beta_g c_{g}^{h} & 2u_g-\beta_g c_{g}^{u} & \beta_g c_{g}^{1} \\
	\sigma_g u_g c_{l}^{h} & -\sigma_g u_g c_{l}^{u} & \sigma_g u_g c_{l}^{1} & -u_g H_g + u_g\beta_g c_{g}^{h} & H_g -u_g\beta_g c_{g}^{u} & u_g+u_g\beta_g c_{g}^{1} \\
	\end{pmatrix}$}
\end{equation}
where
\begin{subequations}\begin{align}
	c_l^h \equiv a_l^2 + \normp{\gamma_l -1}\normp{u_l^2 - H_l} &;\quad c_g^h \equiv a_g^2 + \normp{\gamma_g -1}\normp{u_g^2 - H_g}\\
	c_l^u \equiv \normp{\gamma_l -1}u_l &;\quad c_g^u \equiv \normp{\gamma_g -1}u_g \\
	c_l^1 \equiv \gamma_l -1 &;\quad c_g^1 \equiv \gamma_g -1 \\
	\beta_{l} \equiv \frac{1+\alpha_l\varepsilon_g}{1 + \alpha_g\varepsilon_l + \alpha_l\varepsilon_g} &;\quad \beta_{g} \equiv \frac{1+\alpha_g\varepsilon_l}{1 + \alpha_g\varepsilon_l + \alpha_l\varepsilon_g} \\
	\sigma_{l} \equiv \frac{\alpha_l \varepsilon_l}{1 + \alpha_g\varepsilon_l + \alpha_l\varepsilon_g} &;\quad \sigma_{g} \equiv \frac{\alpha_g \varepsilon_g}{1 + \alpha_g\varepsilon_l + \alpha_l\varepsilon_g}
	\end{align}\end{subequations}
\begin{equation}\label{Eq-A.15}
a_k^2 \equiv \frac{1}{\normp{\frac{\partial \rho_k}{\partial p}}_{h_k} + \frac{1}{\rho_k}\normp{\frac{\partial \rho_k}{\partial h_k}}_{p}}, \quad \gamma_k \equiv \frac{\normp{\frac{\partial \rho_k}{\partial p}}_{h_k}}{\normp{\frac{\partial \rho_k}{\partial p}}_{h_k} + \frac{1}{\rho_k}\normp{\frac{\partial \rho_k}{\partial h_k}}_{p}},\quad \varepsilon_k = \frac{\rho_k a_k^2 - \gamma_k p}{p}
\end{equation}
The approximate eigenvalues and right eigenvectors are
\begin{subequations}\label{theory:refEq-B2}\begin{align}
	\lambda_{1} \approx u_l - \sqrt{\beta_l}a_l; \lambda_{2} &= u_l ; \lambda_{3} \approx u_l + \sqrt{\beta_l}a_l \\
	\lambda_{4} \approx u_g - \sqrt{\beta_g}a_g; \lambda_{5} &= u_g ; \lambda_{6} \approx u_g + \sqrt{\beta_g}a_g
	\end{align}\end{subequations}
\begin{equation}\label{theory:refEq-B3}\begin{split}
\vect{K}_{1} \approx \begin{pmatrix}
1 \\
u_l -\sqrt{\beta_l}a_l \\
H_l -\sqrt{\beta_l}a_l u_l \\
0 \\
0 \\
0 \\
\end{pmatrix},
\vect{K}_{2} &\approx  \begin{pmatrix}
1 \\
u_l \\
H_l - \gamma_l^{*}a_l^2 \\
0 \\
0 \\
0 \\
\end{pmatrix},
\vect{K}_{3} \approx  \begin{pmatrix}
1 \\
u_l +\sqrt{\beta_l}a_l \\
H_l +\sqrt{\beta_l}a_l u_l \\
0 \\
0 \\
0 \\
\end{pmatrix}\\
\vect{K}_{4} \approx \begin{pmatrix}
q_4\\
q_4\lambda_{4} \\
q_4\normb{H_l - u_l^2 + u_l\lambda_{4}}\\
1 \\
u_g -\sqrt{\beta_g}a_g \\
H_g -\sqrt{\beta_g}a_g u_g \\
\end{pmatrix},
\vect{K}_{5} &\approx  \begin{pmatrix}
0\\
0\\
0\\
1 \\
u_g \\
H_g - \gamma_g^{*}a_g^2 \\
\end{pmatrix},
\vect{K}_{6} \approx  \begin{pmatrix}
q_6\\
q_6\lambda_{6}\\
q_6\normb{H_l - u_l^2 + u_l\lambda_{6}}\\
1 \\
u_g +\sqrt{\beta_g}a_g \\
H_g +\sqrt{\beta_g}a_g u_g \\
\end{pmatrix}
\end{split}\end{equation}
where $\gamma_l^{*}=1/\normp{\gamma_l-1}$ and $\gamma_g^{*}=1/\normp{\gamma_g-1}$. $q_4$ and $q_6$ are two auxiliary variables defined as
\begin{equation}
q_4 \equiv \frac{\sigma_l a_g^2}{\normp{\lambda_{4}-\lambda_{1}}\normp{\lambda_{4}-\lambda_{3}}}; \quad
q_6 \equiv \frac{\sigma_l a_g^2}{\normp{\lambda_{6}-\lambda_{1}}\normp{\lambda_{6}-\lambda_{3}}}
\end{equation}
\section{Coefficient matrices/vectors}
The coefficient matrices, $\mat{A}_0$, $\mat{A}_{1}$, and $\mat{A}_{2}$ in the adjoint equation are
\begin{equation}\label{appendix:refEq-D4}
\mat{A}_{0} = \begin{pmatrix}
-\rho_l & \alpha_l x_{11}  & \alpha_l x_{12} & 0 & 0  & 0   \\
-\rho_l u_l & \alpha_l u_l x_{11} & \alpha_l u_l x_{12} & 0 & \alpha_l\rho_l & 0 \\
-\rho_l E_l-p & \alpha_l (E_l x_{11} + \rho_l y_{11}) & \alpha_l (E_l x_{12} + \rho_l y_{12}) & 0 & \alpha_l\rho_l u_l & 0 \\
\rho_g     & \alpha_g x_{21} & 0 & \alpha_g x_{22} & 0 & 0 \\
\rho_g u_g & \alpha_g u_g x_{21} & 0 & \alpha_g u_g x_{22}& 0 & \alpha_g\rho_g \\
\rho_g E_g+p & \alpha_g (E_g x_{21} + \rho_g y_{21}) & 0 & \alpha_g (E_g x_{22} + \rho_g y_{22}) & 0& \alpha_g\rho_g u_g \\
\end{pmatrix}
\end{equation}
\begin{equation}
\mat{A}_{1} = \scalebox{.9}{$\begin{pmatrix}
	-\rho_l u_l & \alpha_l u_l x_{11}&\alpha_l u_l x_{12}&0&\alpha_l \rho_l&0\\
	- \rho_l u_l^2 &\alpha_l\normp{1 + u_l^2 x_{11}}& \alpha_l u_l^2 x_{22}&0&2\alpha_l\rho_l u_l&0\\
	-\rho_l u_l H_l&\alpha_l u_l\normp{E_l x_{11} + \rho_l y_{11} +1}&\alpha_l u_l\normp{E_l x_{12} + \rho_l y_{12}} & 0&\alpha_l\rho_l\normp{H_l + u_l^2}&0\\
	\rho_g u_g&\alpha_g u_g x_{21}& 0&\alpha_g u_g x_{22}&0& \alpha_g\rho_g \\
	\rho_g u_g^2& \alpha_g\normp{1 + u_g^2 x_{21}}& 0& \alpha_g u_g^2 x_{22}&0&2\alpha_g\rho_g u_g\\
	\rho_g u_g H_g & \alpha_g u_g \normp{E_g x_{21} + \rho_g y_{21} +1} & 0 & \alpha_g u_g \normp{E_g x_{22} + \rho_g y_{22}} & 0 & \alpha_g \rho_g \normp{H_g + u_g^2} \\
	\end{pmatrix}$}
\end{equation}
\begin{equation}
\mat{A}_{2} = \begin{pmatrix}
0 & 0 & 0 & 0 & 0 & 0 \\
\frac{\partial p}{\partial x} & -\frac{\partial \alpha_g}{\partial x} & 0 & 0 & 0 & 0 \\
0 & 0 & 0 & 0 & 0 & 0 \\
0 & 0 & 0 & 0 & 0 & 0 \\
-\frac{\partial p}{\partial x} & \frac{\partial \alpha_g}{\partial x} & 0 & 0 & 0 & 0 \\
0 & 0 & 0 & 0 & 0 & 0 \\
\end{pmatrix} + \begin{pmatrix}
\frac{\partial S_l^c}{\partial \alpha_g} & \frac{\partial S_l^c}{\partial p}& \frac{\partial S_l^c}{\partial T_l} & \frac{\partial S_l^c}{\partial T_g} & \frac{\partial S_l^c}{\partial u_l} & \frac{\partial S_l^c}{\partial u_g} \\
\frac{\partial S_l^m}{\partial \alpha_g} & \frac{\partial S_l^m}{\partial p}& \frac{\partial S_l^m}{\partial T_l} & \frac{\partial S_l^m}{\partial T_g} & \frac{\partial S_l^m}{\partial u_l} & \frac{\partial S_l^m}{\partial u_g} \\
\frac{\partial S_l^e}{\partial \alpha_g} & \frac{\partial S_l^e}{\partial p}& \frac{\partial S_l^e}{\partial T_l} & \frac{\partial S_l^e}{\partial T_g} & \frac{\partial S_l^e}{\partial u_l} & \frac{\partial S_l^e}{\partial u_g} \\
\frac{\partial S_g^c}{\partial \alpha_g} & \frac{\partial S_g^c}{\partial p}& \frac{\partial S_g^c}{\partial T_l} & \frac{\partial S_g^c}{\partial T_g} & \frac{\partial S_g^c}{\partial u_l} & \frac{\partial S_g^c}{\partial u_g} \\
\frac{\partial S_g^m}{\partial \alpha_g} & \frac{\partial S_g^m}{\partial p}& \frac{\partial S_g^m}{\partial T_l} & \frac{\partial S_g^m}{\partial T_g} & \frac{\partial S_g^m}{\partial u_l} & \frac{\partial S_g^m}{\partial u_g} \\
\frac{\partial S_g^e}{\partial \alpha_g} & \frac{\partial S_g^e}{\partial p}& \frac{\partial S_g^e}{\partial T_l} & \frac{\partial S_g^e}{\partial T_g} & \frac{\partial S_g^e}{\partial u_l} & \frac{\partial S_g^e}{\partial u_g} \\
\end{pmatrix}
\end{equation}
where $S_{l}^{c}$, $S_{l}^{m}$, $S_{l}^{e}$, $S_{g}^{c}$, $S_{g}^{m}$, and $S_{g}^{e}$ are the source term in the liquid mass, liquid momentum, liquid energy, gas mass, gas momentum, and gas energy, respectively. And
\begin{subequations}\label{appendix:refEq-D3}
	\begin{align}
	x_{11} &= \left(\frac{\partial \rho_l}{\partial p}\right)_{T_l},  x_{12} = \left(\frac{\partial \rho_l}{\partial T_l}\right)_{p}\\
	x_{21} &= \left(\frac{\partial \rho_g}{\partial p}\right)_{T_g},  x_{22} = \left(\frac{\partial \rho_g}{\partial T_g}\right)_{p}\\
	y_{11} &= \left(\frac{\partial e_l}{\partial p}\right)_{T_l},     y_{12} = \left(\frac{\partial e_l}{\partial T_l}\right)_{p}\\
	y_{21} &= \left(\frac{\partial e_g}{\partial p}\right)_{T_g},     y_{22} = \left(\frac{\partial e_g}{\partial T_g}\right)_{p}
	\end{align}
\end{subequations}
In practice, the partial derivatives related to the source terms are computed numerically using a finite difference approximation. \\

The functions $B_1\normp{\vects{\phi}}$ to $B_6\normp{\vects{\phi}}$ are
\begin{equation}\begin{split}
B_1\normp{\vects{\phi}} &= -\rho_l u_l \phi_1 - \normp{\rho_lu_l^2}\phi_2 - \rho_l u_l H_l\phi_3 + \rho_g u_g \phi_4 + \normp{\rho_g u_g^2 }\phi_5 + \rho_g u_g H_g \phi_6\\
B_2\normp{\vects{\phi}} &= \alpha_l u_l x_{11}\phi_1 + \alpha_l\normp{1 + u_l^2 x_{11}} \phi_2 + \alpha_l u_l \normp{E_l x_{11} + \rho_ly_{11}+1}\phi_3 \\
&+ \alpha_g u_g x_{21} \phi_4 + \alpha_g\normp{1 + u_g^2 x_{21}}\phi_5 + \alpha_g u_g \normp{E_g x_{21} + \rho_g y_{21} +1} \phi_6\\
B_3\normp{\vects{\phi}} &= \alpha_l u_l x_{12}\phi_1 + \alpha_l u_l^2 x_{12} \phi_2 + \alpha_l u_l \normp{E_l x_{12} + \rho_l y_{12}}\phi_3 \\
B_4\normp{\vects{\phi}} &= \alpha_g u_g x_{22}\phi_4 + \alpha_g u_g^2 x_{22} \phi_5 + \alpha_g u_g \normp{E_g x_{22} + \rho_g y_{22}}\phi_6 \\
B_5\normp{\vects{\phi}} &= \alpha_l \rho_l \phi_1 + 2\alpha_l \rho_l u_l \phi_2 + \alpha_l \rho_l \normp{H_l + u_l^2}\phi_3 \\
B_6\normp{\vects{\phi}} &= \alpha_g \rho_g \phi_4 + 2\alpha_g \rho_g u_g \phi_5 + \alpha_g \rho_g \normp{H_g + u_g^2}\phi_3
\end{split}\end{equation}
\bibliographystyle{unsrt}

\end{document}